\documentclass[aps, prl, amssymb, longbibliography, twocolumn, superscriptaddress]{revtex4-1}

\usepackage[german,english]{babel}
\usepackage[utf8]{inputenc}

\usepackage{amsmath}
\usepackage{amsthm}
\usepackage{amssymb}
\usepackage{dcolumn}
\usepackage{comment}
\usepackage{mathtools}

\AtBeginDocument{%
    \newwrite\bibnotes
    \def\bibnotesext{Notes.bib}
    \immediate\openout\bibnotes=\jobname\bibnotesext
    \immediate\write\bibnotes{@CONTROL{REVTEX41Control}}
    \immediate\write\bibnotes{@CONTROL{%
    apsrev41Control,author="08",editor="1",pages="1",title="0",year="1"}}
     \if@filesw
     \immediate\write\@auxout{\string\citation{apsrev41Control}}%
    \fi
}%

\usepackage{graphicx}
\usepackage{dcolumn}
\usepackage{bm}
\usepackage{physics}

\usepackage{dcolumn}

\usepackage[version=4]{mhchem}

\usepackage{tabularx}
\usepackage[normalem]{ulem}
\usepackage[usenames, dvipsnames]{color}

\usepackage{placeins}

\usepackage{listings}
\usepackage{xcolor}
\usepackage{algpseudocode}

\usepackage{colortbl}
\usepackage{diagbox}
\usepackage[version=4]{mhchem}

\usepackage{qcircuit}
\usepackage{algorithm}
\usepackage{algpseudocode}
\floatname{algorithm}{Protocol}

\newtheorem{theorem}{Theorem}

\newtheorem{lemma}{Lemma}

\newtheorem{proposition}{Proposition}
\newtheorem{example}{Example}
\newtheorem{conjecture}{Conjecture}
\newtheorem{remark}{Remark}

\definecolor{myrefcolor}{rgb}{0.067,0.5,0.5}

\usepackage[
    breaklinks,
    pdftex,
    colorlinks=true,
    linkcolor=myrefcolor,
    citecolor=myrefcolor,
    urlcolor=myrefcolor
]{hyperref}

\usepackage{times}

\begin{document}
	
\title{Measurement-driven quantum 
advantages in shallow
circuits}
\author{Chenfeng Cao}
\affiliation{Dahlem Center for Complex Quantum Systems, Freie Universit\"{a}t Berlin, 14195 Berlin, Germany}
\author{Jens Eisert}
\email{jense@zedat.fu-berlin.de}
\affiliation{Dahlem Center for Complex Quantum Systems, Freie Universit\"{a}t Berlin, 14195 Berlin, Germany}
\affiliation{Helmholtz-Zentrum Berlin f{\"u}r Materialien und Energie, 14109 Berlin, Germany}

	\date{\today}
	
\begin{abstract}
Quantum advantage schemes probe the boundary between classically simulatable and classically intractable quantum dynamics. We 
explore the impact of mid-circuit measurements on the computational power of quantum circuits. To this effect, we
focus on quantum sampling and introduce a constant-depth measurement-driven approach for efficiently sampling from a broad class of commuting diagonal quantum circuits and associated structured phase states, previously requiring polynomial-depth unitary circuits. By interleaving mid-circuit measurements with feed-forward in randomized ``fan-out staircases'', our dynamical circuits bypass Lieb–Robinson light-cone constraints, enabling global entanglement with flexible auxiliary qubit usage on bounded-degree lattices (e.g., two-dimensional grids). The generated phase states exhibit random‑matrix statistics and anti‑concentration comparable to fully random architectures. We further demonstrate measurement-driven feature maps that distinguish phases of an extended SSH model from random eigenstates in a quantum machine-learning benchmark (reservoir computing). Technologically, our results harness mid-circuit measurements to realize quantum advantages on bounded‑degree hardware with a favorable topology. Conceptually, they provide complexity‑theoretic support for quantum speedups by mid-circuit measurements.

	\end{abstract}
	\maketitle

A leading paradigm for demonstrating quantum advantage is quantum sampling: implementing a quantum evolution and sampling from its output distribution that is believed to be classically intractable under standard assumptions~\cite{Hangleiter2023Computational}. Notable examples include boson sampling~\cite{Aaronson2011TheComputational,Tillmann2013Experimental,Hamilton2017Gaussian,Lund2017Quantum,Deshpande2022Quantum} and random circuit sampling~\cite{Arute2019Quantum,Bouland2019On,Shepherd2009Temporally,Bremner2011Classical,Bremner2016Average-Case,Bremner2017Achieving,Bluvstein2024Logical,Oszmaniec2022Fermion,Haferkamp2020Closing, Ringbauer2025Verifiable,MindTheGaps}. Paradigmatically, 
these tasks are central to clarifying the precise fine-print of quantum advantage, but they also expose a core tension: the deep, highly-connected circuits demanded conflict with the constraints of noisy, locally-connected hardware. This tension motivates the pursuit of hardware‑efficient routes to quantum advantage.

\emph{Instantaneous-quantum-polynomial-time} (IQP) circuits--commuting diagonal circuits whose output probabilities are complex-temperature partition functions—form a central model~\cite{Shepherd2009Temporally,Bremner2011Classical,Bremner2016Average-Case,Bremner2017Achieving,Bluvstein2024Logical, Paletta2024Robust, Jozsa2024Iqp}. The argument for their classical sampling hardness connects the sampling task to the \#P-hard problem of estimating single output probabilities, as an efficient sampler would imply an efficient estimator (leading to an unlikely collapse of the polynomial hierarchy). This connection requires the distribution to exhibit anti-concentration, where probabilities are not dominated by a few outcomes, precluding trivial classical strategies based on guessing dominant bitstrings. Worst-to-average reductions extend the hardness from contrived instances to typical random circuits used in experiments~\cite{Hangleiter2023Computational}. Random, long-range IQP circuits at polynomial depth satisfy these criteria, providing strong evidence for their hardness~\cite{Bremner2016Average-Case,Bremner2017Achieving}. That said, experimentally realizing and witnessing an IQP sampling advantage faces two key challenges: limited hardware connectivity makes the long-range interactions needed for hardness depth-expensive~\cite{Maslov2024Fast}, and noise that grows with depth renders noisy IQP circuits—even with additional \emph{controlled-NOT} (CX) layers—classically simulable 
beyond 
a constant depth under natural noise models~\cite{Nelson2024Polynomial,Rajakumar2025Polynomial}.

A closely related precedent is the quench-based architecture 
of Bermejo-Vega \emph{et al.}~\cite{Bermejo-Vega2018Architectures}, which realizes constant-depth hard sampling on two-dimensional nearest-neighbor lattices using product-state inputs, short-time evolution under a translation-invariant Ising Hamiltonian, and terminal fixed-basis readout---all without mid-circuit measurements. That scheme yields families of nearest-neighbor, translation-invariant 2-local constant-depth IQP circuits, with the complexity-theoretic route proceeding through a nonadaptive measurement-based quantum computing encoding of deeper logical circuits on an enlarged two-dimensional lattice. Our goal here is complementary: we ask whether mid-circuit measurements and feed-forward can compress a class of dense, long-range IQP-type diagonal circuits to constant depth on bounded-degree hardware, using auxiliary qubits that are measured out and do not appear in the final output register.

In this work, we suggest a different approach using mid-circuit measurements, which not only addresses the depth and connectivity challenges facing standard IQP implementations but is also independently interesting. Basically, in our work, we show that constant-depth quantum circuits are substantially more powerful computationally when equipped with mid-circuit measurements compared to the unitary setting. This regime of \emph{noisy intermediate scale quantum} (NISQ) devices with mid-circuit measurements has been dubbed \emph{NISQ+}~\cite{Yihui}: We present a NISQ+/NISQ computational separation under plausible assumptions, highlighting the computational power of measurements along the way. To put this into context, dynamic circuits with measurements and feed-forward have shown great potential in implementing certain deep quantum circuits and preparing highly entangled states~\cite{PhysRevX.9.031009,Piroli2021Quantum,Lu2022Measurement,Deshpande2024Dynamic,DeCross2023Qubit-Reuse,Fossfeig2023Experimental, Buhrman2024State, Iqbal2024Topological, Zi2025Constant, Watts2025Quantum, Mcginley2024Measurement, Yan2024Variational,Elisa,Baumer2024Efficient,Baumer2024Measurement}. By conditioning operations on measurement outcomes, these circuits create non-local ``shortcuts'' that enable the rapid generation of complex, long-range entanglement at minimal depth~\cite{Lieb1972The, Eisert2006General, Bravyi2006Lieb, Baumer2024Efficient,Baumer2024Measurement}. Inspired by these advances, we introduce a measurement-driven ``fan-out staircase'' architecture that implements the dense, long-range interactions required for IQP computational hardness at constant depth, even on simple \emph{two-dimensional} (2D) hardware. Importantly, we show that the output distributions from these shallow circuits retain key signatures of complexity (such as anti-concentration), providing strong evidence for their classical hardness while significantly relaxing the experimental requirements of depth and connectivity.

Beyond establishing a new route to quantum sampling advantages, our measurement-driven technique provides strongly entangled feature maps for quantum machine learning. We demonstrate its effectiveness in quantum reservoir computing, where a fixed quantum evolution acts as a feature map while only a simple classical readout is trained~\cite{Fujii2017Harnessing,Nakajima2019Boosting}. Even under a realistic noise model, our measurement-based reservoir—which leverages global entanglement created at constant depth—accurately classifies distinct topological phases, surpassing conventional local reservoirs.

The upshot of the present work is as follows. Mid-circuit measurements are known to enable the preparation of symmetry-protected topological states~\cite{Buhrman2024State, Iqbal2024Topological} and long-range entangled states~\cite{Lu2022Measurement, Sang2021Entanglement} that elude constant-depth unitary circuits. They also underpin quantum advantage in certain random shallow Clifford architectures~\cite{Watts2025Quantum}. Extending these insights has proved difficult. Here we provide some of the clearest algorithmic-level evidence that measurements with real-time feed-forward enhance shallow quantum algorithms. We do so by compressing dense IQP circuits to constant depth and demonstrating robust anti-concentration, noise resilience, and a provable expressivity advantage in a quantum machine-learning benchmark.

\begin{figure}[t]
\centering
\includegraphics[width=7.9cm]{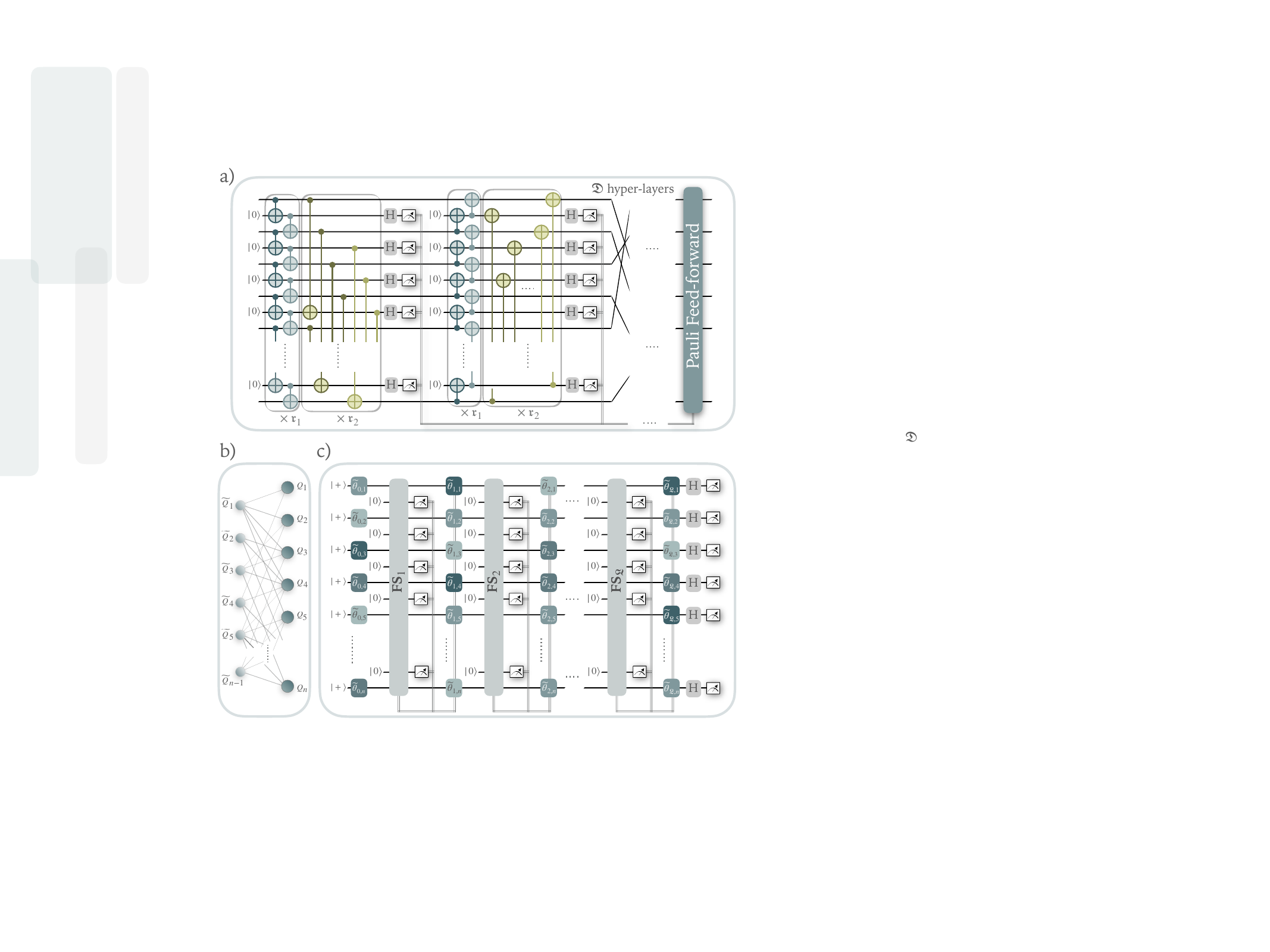}
\caption{Schematic of measurement-based fan-out staircases for dense IQP sampling. (a)~Multi-loop fan-out staircases (alternating-direction ladders) generated by random dynamic circuits with measurement and feed-forward. (b)~Conceptual bipartite (system–auxiliary) coupling graph. Our geometrically local, bounded-degree 2D layouts are subgraphs of this schematic. (c)~Constant-depth implementation of dense IQP sampling via interleaved phase rotations and randomized fan-out staircases, where measurement outcomes inform subsequent corrections. Staircase construction and transfer-matrix feed-forward are detailed in the End Matter and SM.}
\label{fig:Fanout-staircase}
\end{figure}

\paragraph*{Measurement-prepared Hamiltonian phase states.}
The core of our protocol is a constant-depth, measurement-driven method for implementing the dense, many-body Pauli-$Z$ rotations that constitute computationally hard IQP circuits. While one could conventionally generate such a term by conjugating a single-qubit rotation with a deep unitary \(\mathrm{CX}\) circuit, this requires polynomial depth on locally connected hardware.

Our measurement-driven ``fan-out staircase'' (FS) blocks--layered circuits of fan-out gates that build global correlations--achieve this same goal in constant depth. The mechanism of a block involves three conceptual steps. First, we locally entangle system qubits with nearby auxiliary qubits using a short pattern of 
nearest-neighbor CX gates. Next, we measure the auxiliaries in the $X$ basis. An $X$-basis outcome $m_k=1$ is equivalent to having applied a Pauli-$Z$ on that auxiliary qubit just before its measurement. The resulting Pauli-$Z$ correction string on the system qubits is determined by tracking how this ``hypothetical'' $Z$ operator is transformed under conjugation by the full entangling unitary. This transformation—calculated by iteratively applying commutation rules (e.g., \(\mathrm{CX}(c,t)\,Z_t = Z_c Z_t\,\mathrm{CX}(c,t)\)) through the circuit’s layers—establishes a linear map from the measurement-outcome vector $\mathbf m$ to the correction string, captured by a binary transfer matrix \(\mathcal{T}\). The exact FS construction is characterized by a depth parameter \(\mathfrak{D}\) and connectivity parameters \(\mathfrak{r}_1,\mathfrak{r}_2\), as detailed in the End Matter (Protocol~\ref{alg:staircase}) and illustrated in Fig.~\ref{fig:Fanout-staircase}(a).

By interleaving $\mathfrak{L}$ layers of FS blocks and single-$Z$ rotations, the full circuit is
\begin{equation}
\mathcal{C}_{\mathrm{MD}}
= \bigotimes_{j=1}^n e^{\mathrm{i}\vartheta_{\mathfrak{L}+1,j} Z_j}
\left(\prod_{i=1}^{\mathfrak{L}} \mathsf{FS}_i \; \bigotimes_{j=1}^n e^{\mathrm{i}\vartheta_{i,j} Z_j}\right),
\end{equation}
We absorb the Pauli‑frame corrections by classically updating the subsequent rotation angles via
\(\tilde{\vartheta}_{i+1,j} \coloneqq \vartheta_{i+1,j}
+ \frac{\pi}{2}\,(\mathcal{T}^{(i)}\mathbf{m}^{(i)})_j \pmod{2\pi}\), as depicted in Fig.~\ref{fig:Fanout-staircase}(c). Since an FS block conjugates single-qubit $Z$'s into $Z$-strings, the constant-depth protocol effectively realizes an IQP circuit \(\mathcal{C}_{\mathbf{A}, \boldsymbol{\vartheta}} = \exp\left(\mathrm{i}\sum_{i=1}^{s}\vartheta_i\bigotimes_{j=1}^{n}Z^{\mathbf A_{i,j}}\right)\) and prepares the Hamiltonian phase state vectors \(|\psi_{\mathbf A,\boldsymbol\vartheta}\rangle=\mathcal C_{\mathbf A,\boldsymbol\vartheta}|+\rangle^{\otimes n}\). These are defined by a binary matrix \(\mathbf A\in\{0,1\}^{s\times n}\), hereafter termed the \emph{architecture matrix}, 
which our protocol constructs row by row from the Pauli-$Z$ string generated by FS blocks, generating random IQP architectures exhibiting anti-concentration.

\begin{proposition}[Anti-concentration]\label{prop:anti-concentration}
Fully random architectures \(\mathbf{A} \in \mathbb{Z}_2^{s \times n}\) achieve anti-concentration with \(s\in\mathcal O(n)\).
\end{proposition}

\noindent Restricting the angles leads to our first main result.

\begin{theorem}[Measurement-driven short IQP circuits]\label{thm:md-iqp}
Fix $k\ge 1$. For angles $\boldsymbol{\vartheta}\in\{\ell\pi/2^{k}\}_{\ell=0}^{2^{k}-1}$,
the measurement-driven circuit $\mathcal{C}_{\mathrm{MD}}$ implements a dense $k$-local IQP circuit.
\end{theorem}
\noindent Here, a \emph{$k$-local} IQP circuit consists of commuting $Z$-interactions involving at most $k$ qubits (i.e., weight $\le k$).
We call it \emph{dense} if an $n$-independent constant fraction of admissible $Z$-strings appear with \(\Theta(1)\) angles. In particular, the diagonal gates $e^{\mathrm{i}\pi Z/8}$ and $e^{\mathrm{i}\pi ZZ/4}$ arise from the synthesis with $k=3$ and $k=2$, respectively, using a constant number of FS blocks on bounded-degree 2D layouts. To quantitatively evaluate the statistical randomness of binary circuit architectures, we propose the following criterion:

\begin{figure}[t]
\centering
\includegraphics[width=8.65cm]{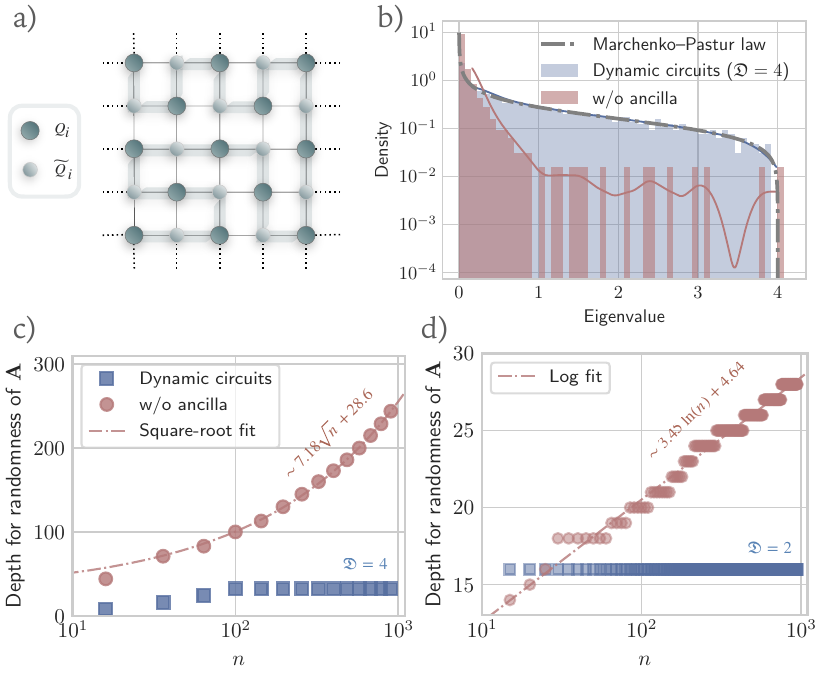}
\caption{(a)~Checkerboard layout of alternating system and auxiliary qubits on a 2D square lattice, with a sampled Hamiltonian path covering all qubits. (b)~Comparison of eigenvalue distributions from randomized fan-out staircases with 
the theoretical Marchenko–Pastur law for a representative $41 \times 41$-qubit system. (c, d)~Minimal circuit depths required to satisfy Criterion 1, contrasting measurement-driven fan-out staircases (blue squares) 
with circuits without auxiliary qubits (red circles), for (c) 2D grid and (d) all-to-all connectivity.}
\label{fig:Criterion}
\end{figure}

\textit{Criterion 1:} $\mathbf{A}\in \mathbb{Z}_2^{s \times n}$ is statistically random if (i) the eigenvalue distribution of the standardized covariance matrix \((2 \mathbf{A}-1)(2 \mathbf{A}-1)^{\top}/s\) matches Marchenko–Pastur law~\cite{Marchenko1967Distribution}; (ii) row and column Hamming weights and pairwise distances follow binomial distributions; and (iii) $\geq 90\%$ of random submatrices exhibit ranks within 2 of full rank over \text{GF(2)}.

\noindent \emph{Remark.} Our randomness diagnostics probe typicality only---analogous to statistical validation in boson and random-circuit sampling---and do not imply a worst-to-average reduction~\cite{Hangleiter2023Computational}. Such a reduction would require distribution-specific random self-reducibility, which we leave open.

These conditions guarantee spectral, statistical, and algebraic universality~\cite{Marchenko1967Distribution,Edelman2005Random,Tao2012Topics}. Using Criterion 1, we numerically compare measurement-driven randomized FS against random \(\mathrm{CX}\) circuits without auxiliary qubits under both 2D grid and all-to-all connectivity. In the 2D configuration, system and auxiliary qubits alternate, with random directed Hamiltonian paths generated according to Protocol 2 in the End Matter, respecting local constraints [Fig.~\ref{fig:Criterion}(a)]. Fig.~\ref{fig:Criterion}(b) shows eigenvalue distributions from measurement-driven FS closely follow the Marchenko–Pastur law, unlike random local \(\mathrm{CX}\) circuits at comparable depths. Numerical results [Fig.~\ref{fig:Criterion}(c,d)] indicate that circuits without auxiliary qubits require polynomial (2D grid) or logarithmic (all-to-all) depths to satisfy Criterion 1. In contrast, measurement-driven FS meet this criterion at constant depth, with parameters $\mathfrak{r}_1=\mathfrak{r}_2=1$, and \(\mathfrak{D}=2\) for all-to-all connectivity and \(\mathfrak{D}=4\) for 2D grid connectivity, even in the large-scale regime. Subsequent IQP simulations employ the 2D grid connectivity, interleaving single-qubit $Z$ rotations with two FS layers (each with \(\mathfrak{D}=2\)).

\begin{figure}[t]
\centering
\includegraphics[width=8.7cm]{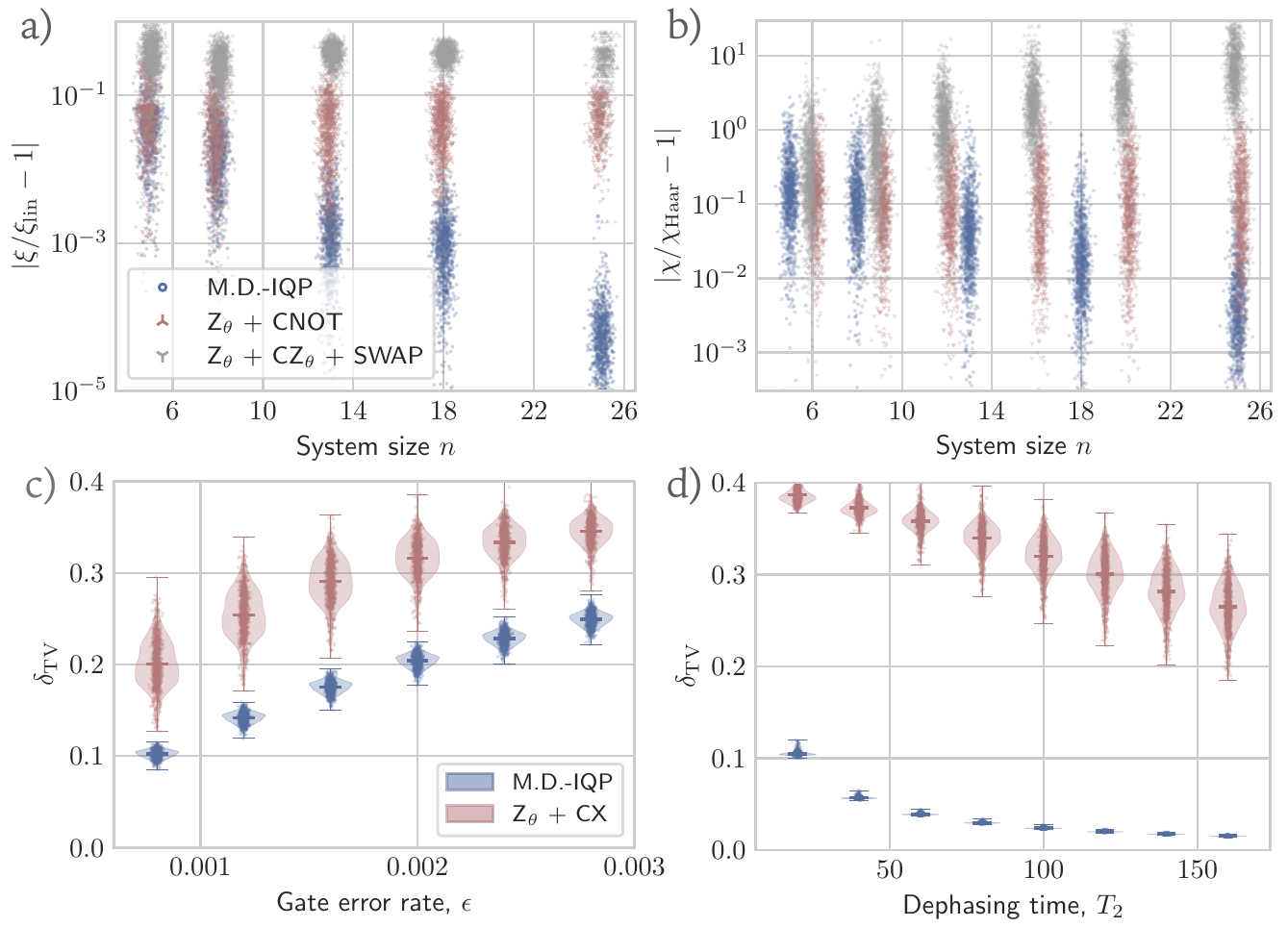}
\caption{Comparison of measurement-driven IQP sampling (blue) with standard IQP sampling (red/gray) at matched constant two-qubit depth on a 2D nearest-neighbor grid. (a)~Relative entanglement-based circuit cost \(\xi/\xi_{\mathrm{lin}}\) for fixed-depth circuits \(\mathfrak{L}=2, \mathfrak{D}=2\) versus system size. (b)~Collision probability ratio \(\chi/\chi_{\text{Haar}}\) versus system size. (c) Total variation distance between noisy and ideal output distributions under gate depolarizing errors. (d) Total variation distance under pure dephasing noise assuming fixed gate duration.}
\label{fig:CP}
\end{figure}

Quantum circuit cost quantifies the minimal resources required to implement a given unitary operation by integrating contributions from local gate interactions, where the entanglement entropy generated across suitable 1D bipartitions provides a fundamental lower bound for this cost~\cite{Marien2016Entanglement, Eisert2021Entangling}. We extend this to 2D lattices by considering horizontal \((\ell_x)\) and vertical \((\ell_y)\) cuts, yielding the summed entanglement entropy
\begin{equation}
\xi(\psi):= \frac{1}{\eta}\biggl(\sum_{\ell_x} S(\psi, \ell_x) + \sum_{\ell_y} S(\psi, \ell_y)\biggr),
\end{equation}
where \(\eta\) is a geometry-dependent constant. \(\xi\) provides a lower bound for the geometrically local circuit cost required to prepare a given state from a product in two dimensions (see the SM). We benchmark against \(\xi_{\mathrm{lin}}\), the average for output states of deep, linear-depth random CX circuits (\(\sim 6n\)). Although such constant-depth circuits on 2D lattices are not expected to achieve genuine volume-law entropy across each cut—particularly for larger systems—Fig.~\ref{fig:CP}(a) shows that measurement-driven IQP circuits nonetheless achieve \(\xi/\xi_{\mathrm{lin}} \approx 1\), matching the entangling power of linear-depth CX circuits. In contrast, conventional depth-matched IQP circuits without auxiliary systems fall significantly short.

The collision probability, \(\chi:={\mathbb{E}}_\mathcal{C}\left[\sum_{x} p_\mathcal{C}(x)^2\right]\), quantifies quantum circuit anti-concentration. For sufficiently deep Haar-random circuits, the collision probability approaches $\chi_{\text{Haar}}=2/(2^n+1)$. Anti-concentration requires the ratio \(\chi/\chi_{\text{Haar}}\) to remain bounded by a small constant~\cite{Dalzell2022Random}, a condition underpinning complexity arguments and benchmarks like cross-entropy benchmarking~\cite{Boixo2018Characterizing, Arute2019Quantum}. Fig.~\ref{fig:CP}(b) compares collision probabilities between measurement-driven IQP circuits and standard IQP circuits without auxiliary qubits at comparable depths. Unlike standard IQP circuits, measurement-driven IQP circuits exhibit a decreasing collision probability ratio with increasing system size, indicating robust anti-concentration at constant depth and minimal sensitivity to connectivity constraints.

Building upon (i) Prop.~\ref{prop:anti-concentration}, (ii) our 2D numerics, and (iii) Lemma~8 of Ref.~\cite{Hangleiter2018Anticoncentration}, which establishes \#P-hardness for depth-$\mathcal{O}(n)$ random circuits over $\{e^{\mathrm{i}\pi Z/8},\,e^{\mathrm{i}\pi ZZ/4}, \mathrm{SWAP}\}$, we make the following observation. On bounded-degree 2D layouts with an interleaved system–auxiliary placement, our constant two-qubit depth protocol implements a subclass of dense $k$-local diagonal layers, including the gates $e^{\mathrm{i}\pi Z/8}$ and $e^{\mathrm{i}\pi ZZ/4}$ central to that hardness result (Thm.~\ref{thm:md-iqp}). Since the SWAP permutations also required by the full construction in Ref.~\cite{Hangleiter2018Anticoncentration} are realizable in our model via measurement-mediated routing without added unitary depth, our scheme therefore realizes a subset of instances from this known-hard circuit family at constant depth. This serves as strong motivation—not a formal worst-to-average reduction~\cite{Bouland2019On,Haferkamp2020Closing,Bremner2016Average-Case}—for the more permissive conjecture that follows.

\begin{conjecture}[Average-case hardness]\label{conj:core-hardness}
For a broad class of connected bounded‑degree bipartite lattices in dimension \(d\ge2\),
there exists a constant‑two‑qubit‑depth, measurement‑driven IQP family—realized via
randomized fan‑out staircases on \(n\) qubits with at most \(O(n\log n)\) auxiliaries and
i.i.d.\ phases in \([0,2\pi)\)—whose output distributions anti‑concentrate and for which,
on a non‑negligible fraction of instances, at least one fixed output probability is
\#P‑hard to approximate within a constant multiplicative factor.
\end{conjecture}

\noindent\emph{Remark.} Our numerics suggest \(\mathcal O(n)\) auxiliaries often suffice on 2D grids; the \(\mathcal O(n\log n)\) budget in the conjecture is a conservative high‑probability allowance to guarantee strong pseudo‑randomness across a wider range of bounded‑degree layouts (e.g., by modestly widening fan‑out targets or layering additional auxiliary resources), at constant two‑qubit depth.

\noindent\emph{Implication.} Assuming Conjecture~\ref{conj:core-hardness} and non-collapse of the polynomial hierarchy, the standard Stockmeyer route implies that approximate sampling from this family is classically intractable up to inverse-polynomial total-variation error~\cite{Hangleiter2023Computational, Stockmeyer1983}.

Measurement-driven IQP circuits exhibit notably enhanced noise resilience compared to conventional IQP circuits, which typically become classically simulable under constant gate noise beyond certain depth thresholds~\cite{Nelson2024Polynomial,Rajakumar2025Polynomial}. By compressing a family of polynomial-depth circuits into constant depth, measurement-driven circuits evade these simulation methods typically applicable to deeper noisy circuits.

To quantify this advantage, we numerically evaluate the total variation distance $\delta_\mathrm{TV}(p,\widetilde{p}) := \sum_{x}\left|p(x)-\widetilde{p}(x)\right|/2$ between the ideal output distribution $p(x)$ and the noisy distribution $\widetilde{p}(x)$ for standard and measurement-driven IQP circuits on a \(6 \times 6\) 2D grid, assuming ideal measurements. We first analyze gate depolarizing noise at varying error rates (Fig.~\ref{fig:CP}(c)). Measurement-driven circuits consistently yield lower \(\delta_\mathrm{TV}\) compared to standard IQP circuits, primarily due to efficient long-range interactions without additional \(\mathrm{SWAP}\) gates. Moreover, measurement-driven circuits exhibit narrower \(\delta_\mathrm{TV}\) distributions, as local errors propagate into correlated global errors through feed-forward corrections.

In a second analysis, we omit gate errors and focus exclusively on decoherence characterized by the dephasing coherence time $T_2$ (assuming an infinite $T_1$). The total variation distance $\delta_{\mathrm{TV}}$ is expected to saturate exponentially with circuit duration $\Delta t$ according to: \(
\delta_{\mathrm{TV}}(\Delta t) \approx \delta_{\mathrm{TV},\infty} (1- e^{-\kappa \Delta t})\),
where \(\delta_{\mathrm{TV},\infty}\) is the asymptotic distance between the ideal and uniform distributions, and $\kappa$ is an effective dephasing rate. Numerical results (assuming 200 ns per \(\mathrm{CX}\) layer) in Fig.~\ref{fig:CP}(d) show that depth compression in measurement-driven circuits significantly reduces \(\delta_{\mathrm{TV}}\), demonstrating substantial mitigation of decoherence effects.

\paragraph*{Measurement-based quantum reservoir.} \emph{Reservoir computing} (RC) embeds inputs using a fixed dynamical system while training only a simple classical readout~\cite{Jaeger2001Echo, Maass2002Realtime}. In quantum RC, the reservoir is a fixed quantum evolution whose intrinsic dynamics generate rich features, with measurements inducing nonlinearity and entanglement enabling long-range mixing~\cite{Fujii2017Harnessing,Nakajima2019Boosting}. In our scheme, interleaving single-qubit gates with $\mathsf{FS}$ and $\mathsf{FS}^\dagger$ forms a constant-depth reservoir where mid-circuit measurements and Pauli-frame feed-forward create input-dependent nonlinear feature maps.

We benchmark our measurement-based quantum reservoir to an extended bosonic \emph{Su–Schrieffer–Heeger chain}, $\mathcal{H} = \sum_{\mu \in \{x, y, z\}} \gamma_\mu \left( J \sum_{i=1}^{n / 2} \sigma_{2i-1}^\mu \sigma_{2i}^\mu + J' \sum_{i=1}^{n / 2 - 1} \sigma_{2i}^\mu \sigma_{2i+1}^\mu \right)$, where \(\gamma_\mu = 1 + (\delta - 1)
\delta_{\mu , z}\). This model features three distinct phases—trivial, topological, and symmetry-broken—depending on the values of $J$, $J'$, and $\delta$~\cite{Su1979Solitons, Leseleuc2019Observation, Elben2020Many,Cao2025Unveiling}. For each phase, we sample 500 eigenstates uniformly from the lowest 20 levels, apply weak local perturbations, and evolve under the candidate reservoirs. Only local spin observables \(\langle \sigma^z_i \rangle\) are measured for each reservoir (8192 shots), with a simulated symmetric per‑qubit readout error of \(0.5\%\).

We compare four reservoirs: (i) a random anisotropic Heisenberg reservoir, $\mathcal{H}_{\mathrm{H}} = \sum_{\langle i,j \rangle}\sum_{\mu} J_{i,j}^{\mu} \, \sigma^{\mu}_i \sigma^{\mu}_j$, where \(\mu \in \{x, y, z\}\) labels spin components;
(ii) a random transverse-field Ising reservoir, $\mathcal H_{\mathrm{TFI}} = \sum_{\langle i,j\rangle} J_{i,j} \sigma^z_i \sigma^z_j + \sum_i h_{i} \sigma^x_i$; (iii) a random transverse-field XY reservoir $\mathcal H_{\mathrm{XY}} = \sum_{\langle i,j\rangle} (J^x_{i,j} \sigma^x_i \sigma^x_j + J^y_{i,j} \sigma^y_i \sigma^y_j) + \sum_i h_{i} \sigma^x_i$; and (iv) a measurement-based multibody XY reservoir $\mathcal{H}_{\mathbf{A}} = \sum_{i=1}^s (c^x_i\prod_{j \in A_i} \sigma^x_j + c^y_i\prod_{j \in A_i} \sigma^y_j )$, with all coupling coefficients $J_{i,j}^{x,y,z}$, $h_{i}$, and $c_i^{x,y,z}$ drawn i.i.d. from $\mathcal{N}(0, 1/2)$, $s = 16$, and \(\mathbf{A}\) encodes the measurement‑enabled architecture. All reservoirs use 16 system qubits on 2D-square and hexagonal lattices, undergo ten identical Floquet cycles generated by their Hamiltonian terms. The classification of phases is performed using a support vector machine (SVM) with a radial basis function kernel~\cite{Cortes1995Support, Chang2011LIBSVM}.

\begin{figure}[t]
\centering
\includegraphics[width=7.8cm]{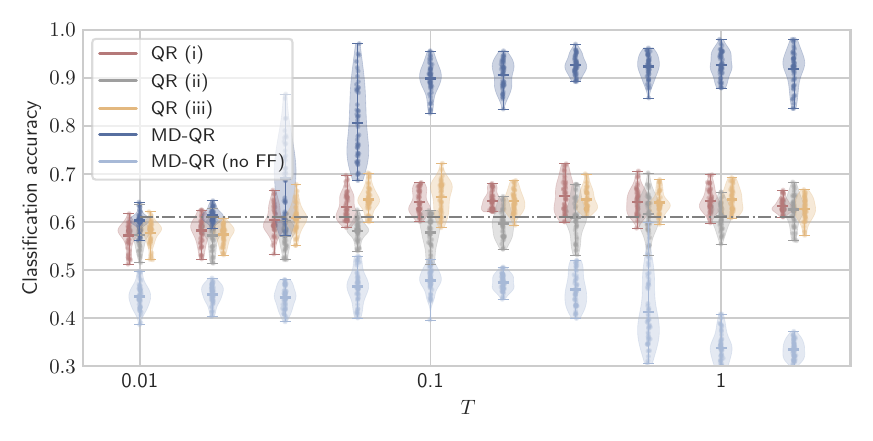}
\caption{Phase-classification accuracy versus Floquet cycle for different reservoirs. Each violin distribution is obtained from 50 random realizations of architectures and coupling strengths. The measurement-driven multibody XY reservoir achieves high accuracy at short times, outperforming local reservoirs (i)-(iii), while removing \emph{feed-forward} (FF) adaptivity significantly reduces performance.}\label{fig:Reservoir}
\end{figure}

The classification accuracy versus Floquet cycle for the reservoirs is shown in Fig.~\ref{fig:Reservoir}, with each violin distribution obtained from 50 independent random realizations of \(\mathbf{A}\) and coupling strengths. In contrast to local reservoirs (i)–(iii), the measurement‑driven multibody reservoir rapidly reaches high accuracy at short times. However, removing the \emph{feed‑forward} (FF) adaptivity leads to a marked drop in performance. From an expressivity standpoint, local-Hamiltonian reservoirs with constant Floquet cycles cannot approximate certain measurement-based feature maps even within inverse-polynomial error, a distinction formalized in Thm.~\ref{thm:reservoir-separation} (proof and additional benchmarks in the SM).

\begin{theorem}[Expressivity separation for local Hamiltonian reservoirs]\label{thm:reservoir-separation} Let $G$ be a connected bounded‑degree graph with $n$ qubits and
fix a Floquet period $T$.
There exist encoded input state vectors $|\psi_0\rangle,|\psi_1\rangle$, a
two‑outcome observable $O$, and a measurement‑driven Floquet reservoir
$\mathcal R_{\mathrm{MD}}^{(T)}$ such that
\begin{equation}
\bigl\lvert
\Tr\bigl[O\,\mathcal{R}_{\mathrm{MD}}^{(T)}(\lvert\psi_0\rangle)\bigr]
-\Tr\bigl[O\,\mathcal{R}_{\mathrm{MD}}^{(T)}(\lvert\psi_1\rangle)\bigr]
\bigr\rvert = \Omega(1),
\end{equation}
while any nearest-neighbor Hamiltonian-based Floquet evolution on the same graph $G$, with the same total evolution time, achieves a separation at most exponentially small in $n$. 
\end{theorem}

\paragraph*{Discussion and outlook.} 
Our measurement-driven framework incorporates extensive interactions between system and auxiliary qubits to broaden the scope of dynamic quantum circuits. In contrast to depth-heavy random-unitary experiments~\cite{Arute2019Quantum}, our dynamic-circuit scheme achieves anti-concentration at constant two-qubit depth via mid-circuit measurements with modest auxiliary overhead. Beyond sampling, its associated Hamiltonian-phase-state preparations can enhance cryptographic protocols~\cite{Bostanci2024Efficient}. An extension to fermion sampling is also promising, as mid-circuit measurements can introduce the non-Gaussian effects crucial for computational hardness~\cite{Oszmaniec2022Fermion}. Furthermore, our measurement-based feature maps can advance quantum machine learning applications by enabling constant-depth global operations that capture data structures inaccessible to local unitary kernels~\cite{Schuld2019Quantum, Havlicek2019Supervised, Gil-Fuster2024On}.

The primitives our protocol requires—mid‑circuit measurements, active reset, and low‑latency feed‑forward—have been demonstrated on leading platforms. Superconducting processors enable real‑time conditional logic via parity measurements and fast (sub‑$\mu$s–few‑$\mu$s) resets~\cite{Riste2013Deterministic, Magnard2018Fast, Krinner2022Realizing, Google2023Suppressing, Chen2025Nishimori}. Trapped‑ion systems, benefiting from long coherence times, support conditional operations in modular QCCD architectures and through networked feed‑forward~\cite{Schindler2011Experimental, Pino2021Demonstration, RyanAnderson2024High, Fossfeig2023Experimental}, while neutral‑atom arrays have achieved site‑selective conditional branching~\cite{Graham2023Midcircuit, Deist2022Mid-Circuit}. These milestones confirm the availability of the necessary NISQ+ primitives. Our classical post‑processing, a sparse Pauli‑frame parity update, adds negligible latency overhead compared to typical gate times and coherence windows.

Nevertheless, mid‑circuit measurement errors (below 1\% in leading systems) and reset infidelity, measurement‑induced crosstalk, and latency‑induced idle dephasing remain practical constraints; these effects have been directly characterized across platforms~\cite{Hothem2025Measuring, Motlakunta2024Preserving, Govia2023Arandomized}. In our constant‑depth realization, each hyper‑layer executes a global Pauli‑frame parity update, so performance is set by the effective mid‑circuit error rate and the feedback delay rather than additional quantum depth. Consistent with Fig.~\ref{fig:CP} and Fig.~\ref{fig:Reservoir}, we expect a graceful degradation under realistic error levels. With sub‑$\mu$s feedback and percent-level readout now available, near-term demonstrations on moderate-size dynamic-circuit hardware appear feasible, while scalability will hinge on continued improvements in measurement fidelity and latency.

Regarding quantum error correction, our shallow-circuit techniques offer a promising route to enhance quantum coding protocols based on low-depth random circuits~\cite{Gullans2021Quantum}. Our approach generalizes dynamic-circuit methods for preparing toric code ground states with long-range topological order~\cite{Verresen2022Efficiently, Fossfeig2023Experimental}, and may enable efficient realization of broader classes of quantum \emph{low-density parity-check} (LDPC) codes~\cite{Breuckmann2021Quantum}, thereby providing a viable blueprint for 
scalable and fault-tolerant quantum-information processing.

\paragraph{Conclusion.}
We introduced an explicit, measurement-based scheme achieving dense IQP sampling at constant depth via randomly generated fan-out staircases, bypassing conventional light-cone constraints to enable global entanglement, robust anti-concentration, and high noise resilience. Additionally, we demonstrated that measurement-based circuits effectively serve as quantum reservoirs, efficiently distinguishing topological phases from random eigenstates. Overall, our results highlight that mid-circuit measurements and feed-forward can significantly enhance quantum algorithm performance at shallow depth---alleviating connectivity constraints and opening pathways for NISQ+ quantum advantages as
next steps to be taken towards achieving \emph{fault-tolerant application-scale} \cite{MindTheGaps} quantum computers.

\vspace{2ex}

\paragraph{Acknowledgments.}
\begin{acknowledgments}
We thank Antonio Anna Mele, Marcel Hinsche, Elies Gil-Fuster, Zhenhuan Liu, Julio Carlos Magdalena De La Fuente, and Xiaoyu He for insightful discussions and valuable feedback. This work has been supported by the BMFTR (DAQC, MuniQC-Atoms), Clusters of Excellence (ML4Q, MATH+), the Munich Quantum Valley, Berlin Quantum, the Quantum Flagship (MILLENION, PASQUANS2), the DFG (CRC 183, SPP 2514),
the European Research Council (DebuQC), 
and the Alexander-von-Humboldt Foundation.
\end{acknowledgments}

\bibliography{bibliography}

@article{Oszmaniec2022Fermion,
  title = {Fermion Sampling: A Robust Quantum Computational Advantage Scheme Using Fermionic Linear Optics and Magic Input States},
  author = {Oszmaniec, Micha\l{} and Dangniam, Ninnat and Morales, Mauro E.S. and Zimbor\'as, Zolt\'an},
  journal = {PRX Quantum},
  volume = {3},
  issue = {2},
  pages = {020328},
  numpages = {54},
  year = {2022},
  month = {May},
  publisher = {American Physical Society},
  doi = {10.1103/PRXQuantum.3.020328},
  url = {https://link.aps.org/doi/10.1103/PRXQuantum.3.020328}
}

@misc{Yihui,
author={Yihui Quek},
note={The term NISQ+ has presumably 
      first been used at the ``Seeking Quantum Advantage''
     Workshop 2024, likely by Yihui Quek (private
      communication).},
year=2024}

@misc{Baumer2024Measurement,
      title={Measurement-Based Long-Range Entangling Gates in Constant Depth}, 
      author={Elisa Baumer and Stefan Woerner},
      year={2024},
      eprint={2408.03064},
      archivePrefix={arXiv},
      optoptprimaryClass={quant-ph},
      url={https://arxiv.org/abs/2408.03064}, 
}

@article{Lu2022Measurement,
  title = {Measurement as a Shortcut to Long-Range Entangled Quantum Matter},
  author = {Lu, Tsung-Cheng and Lessa, Leonardo A. and Kim, Isaac H. and Hsieh, Timothy H.},
  journal = {PRX Quantum},
  volume = {3},
  issue = {4},
  pages = {040337},
  numpages = {22},
  year = {2022},
  month = {Dec},
  publisher = {American Physical Society},
  doi = {10.1103/PRXQuantum.3.040337},
  url = {https://link.aps.org/doi/10.1103/PRXQuantum.3.040337}
}

@misc{Bostanci2024Efficient,
      title={{Efficient quantum pseudorandomness from Hamiltonian phase states}}, 
      author={John Bostanci and Jonas Haferkamp and D. Hangleiter and Alexander Poremba},
      year={2024},
      eprint={2410.08073},
      archivePrefix={arXiv},
      optoptprimaryClass={quant-ph},
      url={https://arxiv.org/abs/2410.08073}, 
}

@misc{Elisa,
year={2025},
      eprint={2504.20832},
      archivePrefix={arXiv},
url={https://arxiv.org/abs/2504.20832},
title={{Approximate quantum Fourier transform in logarithmic depth on a line}},
author={Elisa B{\"a}umer and David Sutter and Stefan Woerner}}

@inbook{Rajakumar2025Polynomial,
author = {Joel Rajakumar and James D. Watson and Yi-Kai Liu},
title = {Polynomial-Time Classical Simulation of Noisy \text{IQP} Circuits with Constant Depth},
booktitle = {Proceedings of the 2025 Annual ACM-SIAM Symposium on Discrete Algorithms (SODA)},
chapter = {},
pages = {1037-1056},
doi = {10.1137/1.9781611978322.30},
URL = {https://epubs.siam.org/doi/abs/10.1137/1.9781611978322.30}
}

@article{Baumer2024Efficient,
  title = {Efficient Long-Range Entanglement Using Dynamic Circuits},
  author = {B\"aumer, Elisa and Tripathi, Vinay and Wang, Derek S. and Rall, Patrick and Chen, Edward H. and Majumder, Swarnadeep and Seif, Alireza and Minev, Zlatko K.},
  journal = {PRX Quantum},
  volume = {5},
  issue = {3},
  pages = {030339},
  numpages = {20},
  year = {2024},
  month = {Aug},
  publisher = {American Physical Society},
  doi = {10.1103/PRXQuantum.5.030339},
  url = {https://link.aps.org/doi/10.1103/PRXQuantum.5.030339}
}

@misc{Jozsa2024Iqp,
      title={\text{IQP} computations with intermediate measurements}, 
      author={Richard Jozsa and Soumik Ghosh and Sergii Strelchuk},
      year={2024},
      eprint={2408.10093},
      archivePrefix={arXiv},
      optoptprimaryClass={quant-ph},
      url={https://arxiv.org/abs/2408.10093}, 
}

@misc{Nelson2024Polynomial,
      title={Polynomial-Time Classical Simulation of Noisy Circuits with Naturally Fault-Tolerant Gates}, 
      author={Jon Nelson and Joel Rajakumar and D. Hangleiter and Michael J. Gullans},
      year={2024},
      eprint={2411.02535},
      archivePrefix={arXiv},
      optoptprimaryClass={quant-ph},
      url={https://arxiv.org/abs/2411.02535}, 
}

@article{Shepherd2009Temporally,
author = {Shepherd, Dan  and Bremner, Michael J. },
title = {Temporally unstructured quantum computation},
journal = {Proc. Roy. Soc. A},
volume = {465},
number = {2105},
pages = {1413-1439},
year = {2009},
doi = {10.1098/rspa.2008.0443},

URL = {https://royalsocietypublishing.org/doi/abs/10.1098/rspa.2008.0443}
}

@article{Bremner2011Classical,
author = {Bremner, Michael J.  and Jozsa, Richard  and Shepherd, Dan J. },
title = {Classical simulation of commuting quantum computations implies collapse of the polynomial hierarchy},
journal = {Proc. Roy. Soc. A},
volume = {467},
number = {2126},
pages = {459-472},
year = {2011},
doi = {10.1098/rspa.2010.0301},

URL = {https://royalsocietypublishing.org/doi/abs/10.1098/rspa.2010.0301}
}

@article{Bremner2017Achieving,
  doi = {10.22331/q-2017-04-25-8},
  url = {https://doi.org/10.22331/q-2017-04-25-8},
  title = {Achieving quantum supremacy with sparse and noisy commuting quantum computations},
  author = {Bremner, Michael J. and Montanaro, Ashley and Shepherd, Dan J.},
  journal = {{Quantum}},
  issn = {2521-327X},
  publisher = {{Verein zur F{\"{o}}rderung des Open Access Publizierens in den Quantenwissenschaften}},
  volume = {1},
  pages = {8},
  month = apr,
  year = {2017}
}

@article{Bremner2016Average-Case,
  title = {Average-Case Complexity Versus Approximate Simulation of Commuting Quantum Computations},
  author = {Bremner, Michael J. and Montanaro, Ashley and Shepherd, Dan J.},
  journal = {Phys. Rev. Lett.},
  volume = {117},
  issue = {8},
  pages = {080501},
  numpages = {5},
  year = {2016},
  month = {Aug},
  publisher = {American Physical Society},
  doi = {10.1103/PhysRevLett.117.080501},
  url = {https://link.aps.org/doi/10.1103/PhysRevLett.117.080501}
}

@article{Paletta2024Robust,
  doi = {10.22331/q-2024-05-06-1337},
  url = {https://doi.org/10.22331/q-2024-05-06-1337},
  title = {Robust sparse \text{IQP} sampling in constant depth},
  author = {Paletta, Louis and Leverrier, Anthony and Sarlette, Alain and Mirrahimi, Mazyar and Vuillot, Christophe},
  journal = {{Quantum}},
  issn = {2521-327X},
  publisher = {{Verein zur F{\"{o}}rderung des Open Access Publizierens in den Quantenwissenschaften}},
  volume = {8},
  pages = {1337},
  month = may,
  year = {2024}
}

@article{Hangleiter2018Anticoncentration,
  doi = {10.22331/q-2018-05-22-65},
  url = {https://doi.org/10.22331/q-2018-05-22-65},
  title = {Anticoncentration theorems for schemes showing a quantum speedup},
  author = {Hangleiter, D. and Bermejo-Vega, J. and Schwarz, M. and Eisert, J. },
  journal = {{Quantum}},
  issn = {2521-327X},
  publisher = {{Verein zur F{\"{o}}rderung des Open Access Publizierens in den Quantenwissenschaften}},
  volume = {2},
  pages = {65},
  month = may,
  year = {2018}
}

@article{Bermejo-Vega2018Architectures,
  title = {Architectures for Quantum Simulation Showing a Quantum Speedup},
  author = {Bermejo-Vega, J. and Hangleiter, D. and Schwarz, M. and Raussendorf, Robert and Eisert, J. },
  journal = {Phys. Rev. X},
  volume = {8},
  issue = {2},
  pages = {021010},
  numpages = {22},
  year = {2018},
  month = {Apr},
  publisher = {American Physical Society},
  doi = {10.1103/PhysRevX.8.021010},
  url = {https://link.aps.org/doi/10.1103/PhysRevX.8.021010}
}

@misc{Deshpande2024Dynamic,
      title={Dynamic parameterized quantum circuits: expressive and barren-plateau free}, 
      author={A. Deshpande and M. Hinsche and S. Najafi and K. Sharma and R. Sweke and C. Zoufal},
      year={2024},
      eprint={2411.05760},
      archivePrefix={arXiv},
      optoptprimaryClass={quant-ph},
      url={https://arxiv.org/abs/2411.05760}, 
}

@article{Eisert2021Entangling,
  title = {Entangling Power and Quantum Circuit Complexity},
  author = {Eisert, J.},
  journal = {Phys. Rev. Lett.},
  volume = {127},
  issue = {2},
  pages = {020501},
  numpages = {5},
  year = {2021},
  month = {Jul},
  publisher = {American Physical Society},
  doi = {10.1103/PhysRevLett.127.020501},
  url = {https://link.aps.org/doi/10.1103/PhysRevLett.127.020501}
}

@InProceedings{Remaud2025Ancilla,
author="Remaud, Maxime
and Vandaele, Vivien",
editor="Gl{\"u}ck, Robert
and Kaarsgaard, Robin",
title="Ancilla-Free Quantum Adder with Sublinear Depth",
booktitle="Reversible Computation",
year="2025",
publisher="Springer Nature Switzerland",
address="Cham",
pages="137--154",
isbn="978-3-031-97063-4"
}

@article{DeCross2023Qubit-Reuse,
  title = {Qubit-Reuse Compilation with Mid-Circuit Measurement and Reset},
  author = {DeCross, Matthew and Chertkov, Eli and Kohagen, Megan and Foss-Feig, Michael},
  journal = {Phys. Rev. X},
  volume = {13},
  issue = {4},
  pages = {041057},
  numpages = {22},
  year = {2023},
  month = {Dec},
  publisher = {American Physical Society},
  doi = {10.1103/PhysRevX.13.041057},
  url = {https://link.aps.org/doi/10.1103/PhysRevX.13.041057}
}

@misc{Fossfeig2023Experimental,
      title={Experimental demonstration of the advantage of adaptive quantum circuits}, 
      author={Michael Foss-Feig and Arkin Tikku and Tsung-Cheng Lu and Karl Mayer and Mohsin Iqbal and Thomas M. Gatterman and Justin A. Gerber and Kevin Gilmore and Dan Gresh and Aaron Hankin and Nathan Hewitt and Chandler V. Horst and Mitchell Matheny and Tanner Mengle and Brian Neyenhuis and Henrik Dreyer and David Hayes and Timothy H. Hsieh and Isaac H. Kim},
      year={2023},
      eprint={2302.03029},
      archivePrefix={arXiv},
      optoptprimaryClass={quant-ph},
      url={https://arxiv.org/abs/2302.03029}, 
}

@misc{Maslov2024Fast,
      title={{Fast classical simulation of Harvard/QuEra \text{IQP} circuits}}, 
      author={Dmitri Maslov and Sergey Bravyi and Felix Tripier and Andrii Maksymov and Joe Latone},
      year={2024},
      eprint={2402.03211},
      archivePrefix={arXiv},
      optoptprimaryClass={quant-ph},
      url={https://arxiv.org/abs/2402.03211}, 
}

@article{Bluvstein2024Logical,
	author = {Bluvstein, Dolev and Evered, Simon J. and Geim, Alexandra A. and Li, Sophie H. and Zhou, Hengyun and Manovitz, Tom and Ebadi, Sepehr and Cain, Madelyn and Kalinowski, Marcin and Hangleiter, D. and Bonilla Ataides, J. Pablo and Maskara, Nishad and Cong, Iris and Gao, Xun and Sales Rodriguez, Pedro and Karolyshyn, Thomas and Semeghini, Giulia and Gullans, Michael J. and Greiner, Markus and Vuleti{\'c}, Vladan and Lukin, Mikhail D.},
	date = {2024/02/01},
	doi = {10.1038/s41586-023-06927-3},
	id = {Bluvstein2024},
	isbn = {1476-4687},
	journal = {Nature},
	number = {7997},
	pages = {58--65},
	title = {Logical quantum processor based on reconfigurable atom arrays},
	url = {https://doi.org/10.1038/s41586-023-06927-3},
	volume = {626},
	year = {2024}}

@article{Hangleiter2023Computational,
  title = {Computational advantage of quantum random sampling},
  author = {Hangleiter, Dominik and Eisert, Jens},
  journal = {Rev. Mod. Phys.},
  volume = {95},
  issue = {3},
  pages = {035001},
  numpages = {82},
  year = {2023},
  month = {Jul},
  publisher = {American Physical Society},
  doi = {10.1103/RevModPhys.95.035001},
  url = {https://link.aps.org/doi/10.1103/RevModPhys.95.035001}
}

@article{Haferkamp2020Closing,
  title = {{Closing gaps of a quantum advantage with short-time Hamiltonian dynamics}},
  author = {Haferkamp, J. and Hangleiter, D. and Bouland, A. and Fefferman, B. and Eisert, J. and Bermejo-Vega, J.},
  journal = {Phys. Rev. Lett.},
  volume = {125},
  issue = {25},
  pages = {250501},
  numpages = {7},
  year = {2020},
  month = {Dec},
  publisher = {American Physical Society},
  doi = {10.1103/PhysRevLett.125.250501},
  url = {https://link.aps.org/doi/10.1103/PhysRevLett.125.250501}
}

@article{Ringbauer2025Verifiable,
	author = {Ringbauer, M. and Hinsche, Marcel and Feldker, Thomas and Faehrmann, Paul K. and Bermejo-Vega, Juani and Edmunds, Claire L. and Postler, Lukas and Stricker, Roman and Marciniak, Christian D. and Meth, Michael and Pogorelov, Ivan and Blatt, Rainer and Schindler, Philipp and Eisert, J.  and Monz, Thomas and Hangleiter, D.},
	date = {2025/01/02},
	doi = {10.1038/s41467-024-55342-3},
	id = {Ringbauer2025},
	isbn = {2041-1723},
	journal = {Nature Comm.},
	number = {1},
	pages = {106},
	title = {Verifiable measurement-based quantum random sampling with trapped ions},
	url = {https://doi.org/10.1038/s41467-024-55342-3},
	volume = {16},
	year = {2025}}

@article{Buhrman2024State,
  doi = {10.22331/q-2024-12-09-1552},
  url = {https://doi.org/10.22331/q-2024-12-09-1552},
  title = {State preparation by shallow circuits using feed forward},
  author = {Buhrman, Harry and Folkertsma, Marten and Loff, Bruno and Neumann, Niels M. P.},
  journal = {{Quantum}},
  issn = {2521-327X},
  publisher = {{Verein zur F{\"{o}}rderung des Open Access Publizierens in den Quantenwissenschaften}},
  volume = {8},
  pages = {1552},
  month = dec,
  year = {2024}
}

@article{Deshpande2022Quantum,
author = {Abhinav Deshpande  and Arthur Mehta  and Trevor Vincent  and Nicolás Quesada  and Marcel Hinsche  and Marios Ioannou  and Lars Madsen  and Jonathan Lavoie  and Haoyu Qi  and J.  Eisert  and D. Hangleiter  and Bill Fefferman  and Ish Dhand },
title = {{Quantum computational advantage via high-dimensional Gaussian boson sampling}},
journal = {Science Adv.},
volume = {8},
number = {1},
pages = {eabi7894},
year = {2022},
doi = {10.1126/sciadv.abi7894},
URL = {https://www.science.org/doi/abs/10.1126/sciadv.abi7894}}

@inproceedings{Aaronson2011TheComputational, author = {Aaronson, Scott and Arkhipov, Alex}, title = {The computational complexity of linear optics}, year = {2011}, isbn = {9781450306911}, publisher = {Association for Computing Machinery}, address = {New York, NY, USA}, url = {https://doi.org/10.1145/1993636.1993682}, doi = {10.1145/1993636.1993682}, booktitle = {Proceedings of the Forty-Third Annual ACM Symposium on Theory of Computing}, pages = {333–342}, numpages = {10}, location = {San Jose, California, USA}, series = {STOC '11} }

@article{Hamilton2017Gaussian,
  title = {Gaussian Boson Sampling},
  author = {Hamilton, Craig S. and Kruse, Regina and Sansoni, Linda and Barkhofen, Sonja and Silberhorn, Christine and Jex, Igor},
  journal = {Phys. Rev. Lett.},
  volume = {119},
  issue = {17},
  pages = {170501},
  numpages = {5},
  year = {2017},
  month = {Oct},
  publisher = {American Physical Society},
  doi = {10.1103/PhysRevLett.119.170501},
  url = {https://link.aps.org/doi/10.1103/PhysRevLett.119.170501}
}

@misc{MindTheGaps,
      title={Mind the gaps: The fraught road to quantum advantage},
author={J. Eisert and J. Preskill}, 
      year={2025},
      eprint={2510.19928},
      archivePrefix={arXiv},
      optoptprimaryClass={quant-ph},
      url={https://arxiv.org/abs/2510.19928} 
}

@article{Arute2019Quantum,
	author = {Arute, Frank and
        others},
	date = {2019/10/01},
	doi = {10.1038/s41586-019-1666-5},
	id = {Arute2019},
	isbn = {1476-4687},
	journal = {Nature},
	number = {7779},
	pages = {505--510},
	title = {Quantum supremacy using a programmable superconducting processor},
	url = {https://doi.org/10.1038/s41586-019-1666-5},
	volume = {574},
	year = {2019}}

@article{Tillmann2013Experimental,
	author = {Tillmann, Max and Daki{\'c}, Borivoje and Heilmann, Ren{\'e} and Nolte, Stefan and Szameit, Alexander and Walther, Philip},
	date = {2013/07/01},
	doi = {10.1038/nphoton.2013.102},
	id = {Tillmann2013},
	isbn = {1749-4893},
	journal = {Nature Phot.},
	number = {7},
	pages = {540--544},
	title = {Experimental boson sampling},
	url = {https://doi.org/10.1038/nphoton.2013.102},
	volume = {7},
	year = {2013}}

@article{Bouland2019On,
	author = {Bouland, Adam and Fefferman, Bill and Nirkhe, Chinmay and Vazirani, Umesh},
	date = {2019/02/01},
	doi = {10.1038/s41567-018-0318-2},
	id = {Bouland2019},
	isbn = {1745-2481},
	journal = {Nature Phys.},
	number = {2},
	pages = {159--163},
	title = {On the complexity and verification of quantum random circuit sampling},
	url = {https://doi.org/10.1038/s41567-018-0318-2},
	volume = {15},
	year = {2019}}

@article{Watts2025Quantum,
  title = {Quantum advantage from measurement-induced entanglement in random shallow circuits},
  author = {Bene Watts, Adam and Gosset, David and Liu, Yinchen and Soleimanifar, Mehdi},
  journal = {PRX Quantum},
  volume = {6},
  issue = {1},
  pages = {010356},
  numpages = {25},
  year = {2025},
  month = {Mar},
  publisher = {American Physical Society},
  doi = {10.1103/PRXQuantum.6.010356},
  url = {https://link.aps.org/doi/10.1103/PRXQuantum.6.010356}
}

@article{Marchenko1967Distribution,
  title={Distribution of eigenvalues for some sets of random matrices},
  author={Marchenko, V. A. and Pastur, Leonid A.},
doi={10.1070/SM1967v001n04ABEH001994},
  journal={Matj. Sb. (NS)},
  volume={72},
  number={114},
  pages={4},
  year={1967}
}

@article{Nakata2014Generating,
doi = {10.1088/1367-2630/16/5/053043},
year = {2014},
month = {may},
publisher = {IOP Publishing},
volume = {16},
number = {5},
pages = {053043},
author = {Nakata, Yoshifumi and Koashi, Masato and Murao, Mio},
title = {Generating a state $t$-design by diagonal quantum circuits},
journal = {New J. Phys.}
}

@article{Edelman2005Random, title={Random matrix theory}, volume={14}, DOI={10.1017/S0962492904000236}, journal={Acta Numerica}, author={Edelman, Alan and Rao, N. Raj}, year={2005}, pages={233–297}}

@book{Tao2012Topics,
  title={Topics in random matrix theory},
  author={Tao, Terence},
  volume={132},
  year={2012},
  publisher={American Mathematical Soc.}
}

@article{Dalzell2022Random,
  title = {Random Quantum Circuits Anticoncentrate in Log Depth},
  author = {Dalzell, Alexander M. and Hunter-Jones, Nicholas and Brand\~ao, Fernando G. S. L.},
  journal = {PRX Quantum},
  volume = {3},
  issue = {1},
  pages = {010333},
  numpages = {43},
  year = {2022},
  month = {Mar},
  publisher = {American Physical Society},
  doi = {10.1103/PRXQuantum.3.010333},
  url = {https://link.aps.org/doi/10.1103/PRXQuantum.3.010333}
}

@Article{Designs,
title={Evenly distributed unitaries: On the structure of unitary designs},
  Author                   = {D.  Gross and K. Audenaert and J. Eisert},
  Journal                  = {J. Math. Phys.},
  Year                     = {2007},
  Pages                    = {052104},
  doi={10.1063/1.2716992},
  Volume                   = {48}
}

@article{Lieb1972The,
	author = {Lieb, Elliott H. and Robinson, Derek W.},
	date = {1972/09/01},
	doi = {10.1007/BF01645779},
	id = {Lieb1972},
	isbn = {1432-0916},
	journal = {Comm. Math. Phys.},
	number = {3},
	pages = {251--257},
	title = {The finite group velocity of quantum spin systems},
	url = {https://doi.org/10.1007/BF01645779},
	volume = {28},
	year = {1972}}

@article{Bravyi2006Lieb,
  title = {{Lieb-Robinson bounds and the generation of correlations and topological quantum order}},
  author = {Bravyi, Sergey and Hastings, Matthew B. and Verstraete, Frank},
  journal = {Phys. Rev. Lett.},
  volume = {97},
  issue = {5},
  pages = {050401},
  numpages = {4},
  year = {2006},
  month = {Jul},
  publisher = {American Physical Society},
  doi = {10.1103/PhysRevLett.97.050401},
  url = {https://link.aps.org/doi/10.1103/PhysRevLett.97.050401}
}

@article{Eisert2006General,
  title = {General Entanglement Scaling Laws from Time Evolution},
  author = {Eisert, Jens  and Osborne, Tobias J.},
  journal = {Phys. Rev. Lett.},
  volume = {97},
  issue = {15},
  pages = {150404},
  numpages = {4},
  year = {2006},
  month = {Oct},
  publisher = {American Physical Society},
  doi = {10.1103/PhysRevLett.97.150404},
  url = {https://link.aps.org/doi/10.1103/PhysRevLett.97.150404}
}

@article{Marien2016Entanglement,
	author = {Mari{\"e}n, Micha{\"e}l and Audenaert, Koenraad M. R. and Van Acoleyen, Karel and Verstraete, Frank},
	date = {2016/08/01},
	doi = {10.1007/s00220-016-2709-5},
	id = {Mari{\"e}n2016},
	isbn = {1432-0916},
	journal = {Comm. Math. Phys.},
	number = {1},
	pages = {35--73},
	title = {Entanglement Rates and the Stability of the Area Law for the Entanglement Entropy},
	url = {https://doi.org/10.1007/s00220-016-2709-5},
	volume = {346},
	year = {2016}}

@misc{Shepherd2010Binary,
      title={Binary Matroids and Quantum Probability Distributions}, 
      author={Dan Shepherd},
      year={2010},
      eprint={1005.1744},
      archivePrefix={arXiv},
      optoptprimaryClass={cs.CC},
      url={https://arxiv.org/abs/1005.1744}, 
}

@misc{Zi2025Constant,
      title={Constant-Depth Quantum Circuits for Arbitrary Quantum State Preparation via Measurement and Feedback}, 
      author={Wei Zi and Junhong Nie and Xiaoming Sun},
      year={2025},
      eprint={2503.16208},
      archivePrefix={arXiv},
      optoptprimaryClass={quant-ph},
      url={https://arxiv.org/abs/2503.16208}, 
}

@book{Kolchin1999Random,
  title={Random Graphs},
  author={Kolchin, Valentin Fedorovich},
  number={53},
  year={1999},
  publisher={Cambridge University Press}
}

@article{Nakata2013Diagonal,
author = {Nakata, Yoshifumi and Murao, Mio},
title = {DIAGONAL-UNITARY 2-DESIGN AND THEIR IMPLEMENTATIONS BY QUANTUM CIRCUITS},
journal = {Int. J. Quant. Inf.},
volume = {11},
number = {07},
pages = {1350062},
year = {2013},
doi = {10.1142/S0219749913500627},

URL = { 
    
        https://doi.org/10.1142/S0219749913500627
},
eprint = { 
    
        https://doi.org/10.1142/S0219749913500627
}
}

@article{Fujii2017Harnessing,
  title = {Harnessing Disordered-Ensemble Quantum Dynamics for Machine Learning},
  author = {Fujii, Keisuke and Nakajima, Kohei},
  journal = {Phys. Rev. Appl.},
  volume = {8},
  issue = {2},
  pages = {024030},
  numpages = {20},
  year = {2017},
  month = {Aug},
  publisher = {American Physical Society},
  doi = {10.1103/PhysRevApplied.8.024030},
  url = {https://link.aps.org/doi/10.1103/PhysRevApplied.8.024030}
}

@article{Nakajima2019Boosting,
  title = {Boosting Computational Power through Spatial Multiplexing in Quantum Reservoir Computing},
  author = {Nakajima, Kohei and Fujii, Keisuke and Negoro, Makoto and Mitarai, Kosuke and Kitagawa, Masahiro},
  journal = {Phys. Rev. Appl.},
  volume = {11},
  issue = {3},
  pages = {034021},
  numpages = {17},
  year = {2019},
  month = {Mar},
  publisher = {American Physical Society},
  doi = {10.1103/PhysRevApplied.11.034021},
  url = {https://link.aps.org/doi/10.1103/PhysRevApplied.11.034021}
}

@article{Su1979Solitons,
  title = {Solitons in Polyacetylene},
  author = {Su, Wu  Pei and Schrieffer, John R. and Heeger, Aan J.},
  journal = {Phys. Rev. Lett.},
  volume = {42},
  issue = {25},
  pages = {1698--1701},
  numpages = {0},
  year = {1979},
  month = {Jun},
  publisher = {American Physical Society},
  doi = {10.1103/PhysRevLett.42.1698},
  url = {https://link.aps.org/doi/10.1103/PhysRevLett.42.1698}
}

@article{PhysRevX.9.031009,
  title = {Measurement-Induced Phase Transitions in the Dynamics of Entanglement},
  author = {Skinner, Brian and Ruhman, Jonathan and Nahum, Adam},
  journal = {Phys. Rev. X},
  volume = {9},
  issue = {3},
  pages = {031009},
  numpages = {21},
  year = {2019},
  publisher = {American Physical Society},
  doi = {10.1103/PhysRevX.9.031009}
}

@article{Leseleuc2019Observation,
author = {Sylvain de Léséleuc  and Vincent Lienhard  and Pascal Scholl  and Daniel Barredo  and Sebastian Weber  and Nicolai Lang  and Hans Peter Büchler  and Thierry Lahaye  and Antoine Browaeys },
title = {Observation of a symmetry-protected topological phase of interacting bosons with Rydberg atoms},
journal = {Science},
volume = {365},
number = {6455},
pages = {775-780},
year = {2019},
doi = {10.1126/science.aav9105},
URL = {https://www.science.org/doi/abs/10.1126/science.aav9105}}

@article{Elben2020Many,
author = {Andreas Elben  and Jinlong Yu  and Guanyu Zhu  and Mohammad Hafezi  and Frank Pollmann  and Peter Zoller  and Benoît Vermersch },
title = {Many-body topological invariants from randomized measurements in synthetic quantum matter},
journal = {Science Adv.},
volume = {6},
number = {15},
pages = {eaaz3666},
year = {2020},
doi = {10.1126/sciadv.aaz3666},
URL = {https://www.science.org/doi/abs/10.1126/sciadv.aaz3666}}

@article{Sang2021Entanglement,
  title = {Entanglement Negativity at Measurement-Induced Criticality},
  author = {Sang, Shengqi and Li, Yaodong and Zhou, Tianci and Chen, Xiao and Hsieh, Timothy H. and Fisher, Matthew P.A.},
  journal = {PRX Quantum},
  volume = {2},
  issue = {3},
  pages = {030313},
  numpages = {23},
  year = {2021},
  month = {Jul},
  publisher = {American Physical Society},
  doi = {10.1103/PRXQuantum.2.030313},
  url = {https://link.aps.org/doi/10.1103/PRXQuantum.2.030313}
}

@article{Havlicek2019Supervised,
	author = {Havl{\'\i}{\v c}ek, Vojt{\v e}ch and C{\'o}rcoles, Antonio D. and Temme, Kristan and Harrow, Aram W. and Kandala, Abhinav and Chow, Jerry M. and Gambetta, Jay M.},
	date = {2019/03/01},
	doi = {10.1038/s41586-019-0980-2},
	id = {Havl{\'\i}{\v c}ek2019},
	isbn = {1476-4687},
	journal = {Nature},
	number = {7747},
	pages = {209--212},
	title = {Supervised learning with quantum-enhanced feature spaces},
	url = {https://doi.org/10.1038/s41586-019-0980-2},
	volume = {567},
	year = {2019}}

@article{Schuld2019Quantum,
  title = {Quantum Machine Learning in Feature Hilbert Spaces},
  author = {Schuld, Maria and Killoran, Nathan},
  journal = {Phys. Rev. Lett.},
  volume = {122},
  issue = {4},
  pages = {040504},
  numpages = {6},
  year = {2019},
  month = {Feb},
  publisher = {American Physical Society},
  doi = {10.1103/PhysRevLett.122.040504},
  url = {https://link.aps.org/doi/10.1103/PhysRevLett.122.040504}
}

@article{Gil-Fuster2024On,
doi = {10.1088/2632-2153/ad2f51},
url = {https://dx.doi.org/10.1088/2632-2153/ad2f51},
year = {2024},
month = {apr},
publisher = {IOP Publishing},
volume = {5},
number = {2},
pages = {025003},
author = {Gil-Fuster, Elies and Eisert, Jens and Dunjko, Vedran},
title = {On the expressivity of embedding quantum kernels},
journal = {Mach. Learn. Sc. Tech.}
}

@article{Gullans2021Quantum,
  title = {Quantum Coding with Low-Depth Random Circuits},
  author = {Gullans, Michael J. and Krastanov, Stefan and Huse, David A. and Jiang, Liang and Flammia, Steven T.},
  journal = {Phys. Rev. X},
  volume = {11},
  issue = {3},
  pages = {031066},
  numpages = {23},
  year = {2021},
  month = {Sep},
  publisher = {American Physical Society},
  doi = {10.1103/PhysRevX.11.031066},
  url = {https://link.aps.org/doi/10.1103/PhysRevX.11.031066}
}

@misc{Verresen2022Efficiently,
      title={{Efficiently preparing Schr\"odinger's cat, fractons and non-Abelian topological order in quantum devices}}, 
      author={Ruben Verresen and Nathanan Tantivasadakarn and Ashvin Vishwanath},
      year={2021},
      eprint={2112.03061},
      archivePrefix={arXiv},
      optprimaryClass={quant-ph},
      url={https://arxiv.org/abs/2112.03061}, 
}

@article{Breuckmann2021Quantum,
  title = {Quantum Low-Density Parity-Check Codes},
  author = {Breuckmann, Nikolas P. and Eberhardt, Jens Niklas},
  journal = {PRX Quantum},
  volume = {2},
  issue = {4},
  pages = {040101},
  numpages = {19},
  year = {2021},
  month = {Oct},
  publisher = {American Physical Society},
  doi = {10.1103/PRXQuantum.2.040101},
  url = {https://link.aps.org/doi/10.1103/PRXQuantum.2.040101}
}

@article{Cortes1995Support,
	author = {Cortes, Corinna and Vapnik, Vladimir},
	date = {1995/09/01},
	doi = {10.1007/BF00994018},
	id = {Cortes1995},
	isbn = {1573-0565},
	journal = {Machine Learning},
	number = {3},
	pages = {273--297},
	title = {Support-vector networks},
	url = {https://doi.org/10.1007/BF00994018},
	volume = {20},
	year = {1995}}

@article{Chang2011LIBSVM,
author = {Chang, Chih-Chung and Lin, Chih-Jen},
title = {LIBSVM: A library for support vector machines},
year = {2011},
issue_date = {April 2011},
publisher = {Association for Computing Machinery},
address = {New York, NY, USA},
volume = {2},
number = {3},
issn = {2157-6904},
url = {https://doi.org/10.1145/1961189.1961199},
doi = {10.1145/1961189.1961199},
journal = {ACM Trans. Intell. Syst. Technol.},
month = may,
articleno = {27},
numpages = {27}
}

@article{Lund2017Quantum,
	author = {Lund, A. P. and Bremner, Michael J. and Ralph, T. C.},
	date = {2017/04/13},
	doi = {10.1038/s41534-017-0018-2},
	id = {Lund2017},
	isbn = {2056-6387},
	journal = {npj Quantum Information},
	number = {1},
	pages = {15},
	title = {Quantum sampling problems, BosonSampling and quantum supremacy},
	url = {https://doi.org/10.1038/s41534-017-0018-2},
	volume = {3},
	year = {2017}}

@article{Iqbal2024Topological,
	author = {Iqbal, Mohsin and Tantivasadakarn, Nathanan and Gatterman, Thomas M. and Gerber, Justin A. and Gilmore, Kevin and Gresh, Dan and Hankin, Aaron and Hewitt, Nathan and Horst, Chandler V. and Matheny, Mitchell and Mengle, Tanner and Neyenhuis, Brian and Vishwanath, Ashvin and Foss-Feig, Michael and Verresen, Ruben and Dreyer, Henrik},
	date = {2024/06/25},
	doi = {10.1038/s42005-024-01698-3},
	id = {Iqbal2024},
	isbn = {2399-3650},
	journal = {Comm. Phys.},
	number = {1},
	pages = {205},
	title = {Topological order from measurements and feed-forward on a trapped ion quantum computer},
	url = {https://doi.org/10.1038/s42005-024-01698-3},
	volume = {7},
	year = {2024}}

@article{Mcginley2024Measurement,
  title = {Measurement-Induced Entanglement and Complexity in Random Constant-Depth 2D Quantum Circuits},
  author = {McGinley, Max and Ho, Wen Wei and Malz, Daniel},
  journal = {Phys. Rev. X},
  volume = {15},
  issue = {2},
  pages = {021059},
  numpages = {35},
  year = {2025},
  month = {May},
  publisher = {American Physical Society},
  doi = {10.1103/PhysRevX.15.021059},
  url = {https://link.aps.org/doi/10.1103/PhysRevX.15.021059}
}

@article{Piroli2021Quantum,
  title = {Quantum Circuits Assisted by Local Operations and Classical Communication: Transformations and Phases of Matter},
  author = {Piroli, Lorenzo and Styliaris, Georgios and Cirac, J. Ignacio},
  journal = {Phys. Rev. Lett.},
  volume = {127},
  issue = {22},
  pages = {220503},
  numpages = {6},
  year = {2021},
  month = {Nov},
  publisher = {American Physical Society},
  doi = {10.1103/PhysRevLett.127.220503},
  url = {https://link.aps.org/doi/10.1103/PhysRevLett.127.220503}
}

@article{Yan2024Variational,
  title = {Variational LOCC-Assisted Quantum Circuits for Long-Range Entangled States},
  author = {Yan, Yuxuan and Ma, Muzhou and Zhou, You and Ma, Xiongfeng},
  journal = {Phys. Rev. Lett.},
  volume = {134},
  issue = {17},
  pages = {170601},
  numpages = {7},
  year = {2025},
  month = {Apr},
  publisher = {American Physical Society},
  doi = {10.1103/PhysRevLett.134.170601},
  url = {https://link.aps.org/doi/10.1103/PhysRevLett.134.170601}
}

@article{Riste2013Deterministic,
	author = {Rist{\`e}, D. and Dukalski, M. and Watson, C. A. and de Lange, G. and Tiggelman, M. J. and Blanter, Ya. M. and Lehnert, K. W. and Schouten, R. N. and DiCarlo, L.},
	date = {2013/10/01},
	doi = {10.1038/nature12513},
	id = {Rist{\`e}2013},
	isbn = {1476-4687},
	journal = {Nature},
	number = {7471},
	pages = {350--354},
	title = {Deterministic entanglement of superconducting qubits by parity measurement and feedback},
	url = {https://doi.org/10.1038/nature12513},
	volume = {502},
	year = {2013}}

@article{Magnard2018Fast,
  title = {Fast and Unconditional All-Microwave Reset of a Superconducting Qubit},
  author = {Magnard, P. and Kurpiers, P. and Royer, B. and Walter, T. and Besse, J.-C. and Gasparinetti, S. and Pechal, M. and Heinsoo, J. and Storz, S. and Blais, A. and Wallraff, A.},
  journal = {Phys. Rev. Lett.},
  volume = {121},
  issue = {6},
  pages = {060502},
  numpages = {6},
  year = {2018},
  month = {Aug},
  publisher = {American Physical Society},
  doi = {10.1103/PhysRevLett.121.060502},
  url = {https://link.aps.org/doi/10.1103/PhysRevLett.121.060502}
}

@article{Pino2021Demonstration,
	author = {Pino, J. M. and Dreiling, J. M. and Figgatt, C. and Gaebler, J. P. and Moses, S. A. and Allman, M. S. and Baldwin, C. H. and Foss-Feig, M. and Hayes, D. and Mayer, K. and Ryan-Anderson, C. and Neyenhuis, B.},
	date = {2021/04/01},
	doi = {10.1038/s41586-021-03318-4},
	id = {Pino2021},
	isbn = {1476-4687},
	journal = {Nature},
	number = {7853},
	pages = {209--213},
	title = {Demonstration of the trapped-ion quantum CCD computer architecture},
	url = {https://doi.org/10.1038/s41586-021-03318-4},
	volume = {592},
	year = {2021}}

@article{Graham2023Midcircuit,
  title = {Midcircuit Measurements on a Single-Species Neutral Alkali Atom Quantum Processor},
  author = {Graham, T. M. and Phuttitarn, L. and Chinnarasu, R. and Song, Y. and Poole, C. and Jooya, K. and Scott, J. and Scott, A. and Eichler, P. and Saffman, M.},
  journal = {Phys. Rev. X},
  volume = {13},
  issue = {4},
  pages = {041051},
  numpages = {22},
  year = {2023},
  month = {Dec},
  publisher = {American Physical Society},
  doi = {10.1103/PhysRevX.13.041051},
  url = {https://link.aps.org/doi/10.1103/PhysRevX.13.041051}
}

@article{Chen2025Nishimori,
	author = {Chen, Edward H. and Zhu, Guo-Yi and Verresen, Ruben and Seif, Alireza and B{\"a}umer, Elisa and Layden, David and Tantivasadakarn, Nathanan and Zhu, Guanyu and Sheldon, Sarah and Vishwanath, Ashvin and Trebst, Simon and Kandala, Abhinav},
	date = {2025/01/01},
	doi = {10.1038/s41567-024-02696-6},
	id = {Chen2025},
	isbn = {1745-2481},
	journal = {Nature Physics},
	number = {1},
	pages = {161--167},
	title = {Nishimori transition across the error threshold for constant-depth quantum circuits},
	url = {https://doi.org/10.1038/s41567-024-02696-6},
	volume = {21},
	year = {2025}}

@article{Krinner2022Realizing,
	author = {Krinner, Sebastian and Lacroix, Nathan and Remm, Ants and Di Paolo, Agustin and Genois, Elie and Leroux, Catherine and Hellings, Christoph and Lazar, Stefania and Swiadek, Francois and Herrmann, Johannes and Norris, Graham J. and Andersen, Christian Kraglund and M{\"u}ller, Markus and Blais, Alexandre and Eichler, Christopher and Wallraff, Andreas},
	date = {2022/05/01},
	doi = {10.1038/s41586-022-04566-8},
	id = {Krinner2022},
	isbn = {1476-4687},
	journal = {Nature},
	number = {7911},
	pages = {669--674},
	title = {Realizing repeated quantum error correction in a distance-three surface code},
	url = {https://doi.org/10.1038/s41586-022-04566-8},
	volume = {605},
	year = {2022}}

@article{Google2023Suppressing,
	author = {Acharya, R. 
        and others},
        optnote={and Aleiner, Igor and Allen, Richard and Andersen, Trond I. and Ansmann, Markus and Arute, Frank and Arya, Kunal and Asfaw, Abraham and Atalaya, Juan and Babbush, Ryan and Bacon, Dave and Bardin, Joseph C. and Basso, Joao and Bengtsson, Andreas and Boixo, Sergio and Bortoli, Gina and Bourassa, Alexandre and Bovaird, Jenna and Brill, Leon and Broughton, Michael and Buckley, Bob B. and Buell, David A. and Burger, Tim and Burkett, Brian and Bushnell, Nicholas and Chen, Yu and Chen, Zijun and Chiaro, Ben and Cogan, Josh and Collins, Roberto and Conner, Paul and Courtney, William and Crook, Alexander L. and Curtin, Ben and Debroy, Dripto M. and Del Toro Barba, Alexander and Demura, Sean and Dunsworth, Andrew and Eppens, Daniel and Erickson, Catherine and Faoro, Lara and Farhi, Edward and Fatemi, Reza and Flores Burgos, Leslie and Forati, Ebrahim and Fowler, Austin G. and Foxen, Brooks and Giang, William and Gidney, Craig and Gilboa, Dar and Giustina, Marissa and Grajales Dau, Alejandro and Gross, Jonathan A. and Habegger, Steve and Hamilton, Michael C. and Harrigan, Matthew P. and Harrington, Sean D. and Higgott, Oscar and Hilton, Jeremy and Hoffmann, Markus and Hong, Sabrina and Huang, Trent and Huff, Ashley and Huggins, William J. and Ioffe, Lev B. and Isakov, Sergei V. and Iveland, Justin and Jeffrey, Evan and Jiang, Zhang and Jones, Cody and Juhas, Pavol and Kafri, Dvir and Kechedzhi, Kostyantyn and Kelly, Julian and Khattar, Tanuj and Khezri, Mostafa and Kieferov{\'a}, M{\'a}ria and Kim, Seon and Kitaev, Alexei and Klimov, Paul V. and Klots, Andrey R. and Korotkov, Alexander N. and Kostritsa, Fedor and Kreikebaum, John Mark and Landhuis, David and Laptev, Pavel and Lau, Kim-Ming and Laws, Lily and Lee, Joonho and Lee, Kenny and Lester, Brian J. and Lill, Alexander and Liu, Wayne and Locharla, Aditya and Lucero, Erik and Malone, Fionn D. and Marshall, Jeffrey and Martin, Orion and McClean, Jarrod R. and McCourt, Trevor and McEwen, Matt and Megrant, Anthony and Meurer Costa, Bernardo and Mi, Xiao and Miao, Kevin C. and Mohseni, Masoud and Montazeri, Shirin and Morvan, Alexis and Mount, Emily and Mruczkiewicz, Wojciech and Naaman, Ofer and Neeley, Matthew and Neill, Charles and Nersisyan, Ani and Neven, Hartmut and Newman, Michael and Ng, Jiun How and Nguyen, Anthony and Nguyen, Murray and Niu, Murphy Yuezhen and O'Brien, Thomas E. and Opremcak, Alex and Platt, John and Petukhov, Andre and Potter, Rebecca and Pryadko, Leonid P. and Quintana, Chris and Roushan, Pedram and Rubin, Nicholas C. and Saei, Negar and Sank, Daniel and Sankaragomathi, Kannan and Satzinger, Kevin J. and Schurkus, Henry F. and Schuster, Christopher and Shearn, Michael J. and Shorter, Aaron and Shvarts, Vladimir and Skruzny, Jindra and Smelyanskiy, Vadim and Smith, W. Clarke and Sterling, George and Strain, Doug and Szalay, Marco and Torres, Alfredo and Vidal, Guifre and Villalonga, Benjamin and Vollgraff Heidweiller, Catherine and White, Theodore and Xing, Cheng and Yao, Z. Jamie and Yeh, Ping and Yoo, Juhwan and Young, Grayson and Zalcman, Adam and Zhang, Yaxing and Zhu, Ningfeng and Google Quantum AI},
	date = {2023/02/01},
	doi = {10.1038/s41586-022-05434-1},
	id = {Acharya2023},
	isbn = {1476-4687},
	journal = {Nature},
	number = {7949},
	pages = {676--681},
	title = {Suppressing quantum errors by scaling a surface code logical qubit},
	url = {https://doi.org/10.1038/s41586-022-05434-1},
	volume = {614},
	year = {2023}}

@article{Hothem2025Measuring,
	author = {Hothem, Daniel and Hines, Jordan and Baldwin, Charles and Gresh, Dan and Blume-Kohout, Robin and Proctor, Timothy},
	date = {2025/07/01},
	doi = {10.1038/s41467-025-60923-x},
	id = {Hothem2025},
	isbn = {2041-1723},
	journal = {Nature Comm.},
	number = {1},
	pages = {5761},
	title = {Measuring error rates of mid-circuit measurements},
	url = {https://doi.org/10.1038/s41467-025-60923-x},
	volume = {16},
	year = {2025}}

@article{
Schindler2011Experimental,
author = {Philipp Schindler  and Julio T. Barreiro  and Thomas Monz  and Volckmar Nebendahl  and Daniel Nigg  and Michael Chwalla  and Markus Hennrich  and Rainer Blatt },
title = {Experimental Repetitive Quantum Error Correction},
journal = {Science},
volume = {332},
number = {6033},
pages = {1059-1061},
year = {2011},
doi = {10.1126/science.1203329},
URL = {https://www.science.org/doi/abs/10.1126/science.1203329},
eprint = {https://www.science.org/doi/pdf/10.1126/science.1203329}}

@article{Deist2022Mid-Circuit,
  title = {Mid-Circuit Cavity Measurement in a Neutral Atom Array},
  author = {Deist, Emma and Lu, Yue-Hui and Ho, Jacquelyn and Pasha, Mary Kate and Zeiher, Johannes and Yan, Zhenjie and Stamper-Kurn, Dan M.},
  journal = {Phys. Rev. Lett.},
  volume = {129},
  issue = {20},
  pages = {203602},
  numpages = {7},
  year = {2022},
  month = {Nov},
  publisher = {American Physical Society},
  doi = {10.1103/PhysRevLett.129.203602},
  url = {https://link.aps.org/doi/10.1103/PhysRevLett.129.203602}
}

@article{Motlakunta2024Preserving,
	author = {Motlakunta, Sainath and Kotibhaskar, Nikhil and Shih, Chung-You and Vogliano, Anthony and McLaren, Darian and Hahn, Lewis and Zhu, Jingwen and Habl{\"u}tzel, Roland and Islam, Rajibul},
	date = {2024/08/03},
	doi = {10.1038/s41467-024-50864-2},
	id = {Motlakunta2024},
	isbn = {2041-1723},
	journal = {Nature Comm.},
	number = {1},
	pages = {6575},
	title = {Preserving a qubit during state-destroying operations on an adjacent qubit at a few micrometers distance},
	url = {https://doi.org/10.1038/s41467-024-50864-2},
	volume = {15},
	year = {2024}}

@article{Govia2023Arandomized,
doi = {10.1088/1367-2630/ad0e19},
url = {https://doi.org/10.1088/1367-2630/ad0e19},
year = {2023},
month = {dec},
publisher = {IOP Publishing},
volume = {25},
number = {12},
pages = {123016},
author = {Govia, L C G and Jurcevic, P and Wood, C J and Kanazawa, N and Merkel, S T and McKay, D C},
title = {A randomized benchmarking suite for mid-circuit measurements},
journal = {New Journal of Physics}
}

@misc{RyanAnderson2024High,
      title={High-fidelity and Fault-tolerant Teleportation of a Logical Qubit using Transversal Gates and Lattice Surgery on a Trapped-ion Quantum Computer}, 
      author={C. Ryan-Anderson and N. C. Brown and C. H. Baldwin and J. M. Dreiling and C. Foltz and J. P. Gaebler and T. M. Gatterman and N. Hewitt and C. Holliman and C. V. Horst and J. Johansen and D. Lucchetti and T. Mengle and M. Matheny and Y. Matsuoka and K. Mayer and M. Mills and S. A. Moses and B. Neyenhuis and J. Pino and P. Siegfried and R. P. Stutz and J. Walker and D. Hayes},
      year={2024},
      eprint={2404.16728},
      archivePrefix={arXiv},
      primaryClass={quant-ph},
      url={https://arxiv.org/abs/2404.16728}, 
}

@article{Boixo2018Characterizing,
	author = {Boixo, Sergio and Isakov, Sergei V. and Smelyanskiy, Vadim N. and Babbush, Ryan and Ding, Nan and Jiang, Zhang and Bremner, Michael J. and Martinis, John M. and Neven, Hartmut},
	date = {2018/06/01},
	doi = {10.1038/s41567-018-0124-x},
	id = {Boixo2018},
	isbn = {1745-2481},
	journal = {Nature Physics},
	number = {6},
	pages = {595--600},
	title = {Characterizing quantum supremacy in near-term devices},
	url = {https://doi.org/10.1038/s41567-018-0124-x},
	volume = {14},
	year = {2018}}

@article{Jaeger2001Echo,
  title={The “echo state” approach to analysing and training recurrent neural networks-with an erratum note},
doi={0.24406/publica-fhg-291111},
  author={Jaeger, Herbert},
  note={Bonn, Germany: German national research center for information technology},
journa={{GMD Tech. Rep.}}, 
  volume={148},
  number={34},
  pages={13},
  year={2001},
  publisher={Bonn}
}

@article{Maass2002Realtime,
  author    = {Wolfgang Maass and Thomas Natschl{\"a}ger and Henry Markram},
  title     = {Real‐time computing without stable states: A new framework for neural computation based on perturbations},
  journal   = {Neur. Comp.},
  volume    = {14},
  number    = {11},
  pages     = {2531--2560},
  year      = {2002},
  doi       = {10.1162/089976602760407955}
}

@inproceedings{Stockmeyer1983, author = {Stockmeyer, Larry}, title = {The complexity of approximate counting}, year = {1983}, isbn = {0897910990}, publisher = {Association for Computing Machinery}, address = {New York, NY, USA}, url = {https://doi.org/10.1145/800061.808740}, doi = {10.1145/800061.808740}, booktitle = {Proceedings of the Fifteenth Annual ACM Symposium on Theory of Computing}, pages = {118–126}, numpages = {9}, series = {STOC '83} }

@article{Cao2025Unveiling,
	author = {Cao, Chenfeng and Gambetta, Filippo Maria and Montanaro, Ashley and Santos, Raul A.},
	date = {2025/06/04},
	doi = {10.1038/s41534-025-01038-5},
	id = {Cao2025},
	isbn = {2056-6387},
	journal = {npj Quantum Information},
	number = {1},
	pages = {93},
	title = {Unveiling quantum phase transitions from traps in variational quantum algorithms},
	url = {https://doi.org/10.1038/s41534-025-01038-5},
	volume = {11},
	year = {2025}}

@CONTROL{REVTEX41Control}

@CONTROL{apsrev41Control,author="08",editor="1",pages="1",title="0",year="1"}

\twocolumngrid
\appendix

\clearpage

\begin{center}
	\textbf{\large End Matter}
\end{center}
\makeatletter

\paragraph*{Protocol details: Measurement-driven fan-out staircases.}
Throughout, we arrange system and auxiliary qubits in an alternating pattern and build randomized fan-out staircase (FS) blocks by interleaving short entangling patterns with mid-circuit measurements; the detailed scheme is given in Protocol~\ref{alg:staircase}. Theoretically, an FS block maps a single-qubit $Z$ phase into a many-body $Z$ rotation via conjugation,
$\mathsf{FS}_i\, e^{\mathrm{i}\vartheta_j Z_j}\, \mathsf{FS}_i^\dagger
= \exp\big(\mathrm{i}\vartheta_j \bigotimes_{\ell=1}^n Z_\ell^{\mathbf A_{i,\ell}}\big)$.
For state preparation from $\lvert +\rangle^{\otimes n}$, this simplifies: $\mathsf{FS}_i^\dagger$ acts trivially. In each FS ladder (forward/backward along a chosen path) we apply nearest-neighbor CXs between systems and auxiliaries, measure all auxiliaries in the $X$ basis, and absorb the induced Pauli-$Z$ string on the systems via Pauli-frame feed-forward using the binary transfer matrix \(\mathcal{T}\).
We denote the forward/backward ladder at round $j$ by \(\mathsf{FL}_j^{(\uparrow/\downarrow)}\) (optionally, \(\mathsf{FL}_j^{(\uparrow)}:=\prod_{i=1}^{n-1}\textsc{FanOut}(Q_{\ell_i}^{(j)}\to T_{\ell_i}^{(j,\uparrow)})\)) and a staircase with \(2\mathfrak D\) measurement rounds by \(\mathsf{FS}:=\prod_{j=0}^{\mathfrak D-1}\big(\mathsf{FL}_{\mathfrak D-j}^{(\downarrow)}\mathsf{FL}_{\mathfrak D-j}^{(\uparrow)}\big)\). A staircase with $2\mathfrak D$ measurement rounds thus has constant two-qubit depth $4\mathfrak D$ (independent of $n$), while the total two-qubit gate count is \(\mathcal O(\mathfrak D n)\).
In this Letter we focus on the minimal-auxiliary setting with $(n-1)$ auxiliaries; however, adding auxiliary–auxiliary couplings extends the scheme to richer interaction patterns without changing the constant-depth scheduling. This dynamic construction circumvents Lieb–Robinson light-cone limits and rapidly generates long-range entanglement on bounded-degree layouts.

\begin{proposition}[Measurement-driven fan-out staircases]
Implementing the randomized fan-out staircase circuits described above without auxiliary qubits necessitates a circuit depth of at least \(\Omega(\log n)\) under all-to-all connectivity. A matching logarithmic depth can be achieved in the simplest scenario, where each fan-out targets a fixed contiguous set of qubits (via Ref.~\cite{Remaud2025Ancilla}).
\end{proposition}

\paragraph{Proof:}
Implementing randomized fan-out staircases without auxiliary qubits demands each input qubit's information to influence $\mathcal{O}(n)$ qubits to achieve global entanglement. Local two-qubit gates impose constraints on information propagation, allowing at most an exponential growth in the range of influence per layer. Thus, after $d$ layers, an input qubit can influence at most $\mathcal{O}(2^d)$ qubits. To ensure global entanglement (influencing $\mathcal{O}(n)$ qubits), the following 
\begin{equation}
2^d \geq \mathcal{O}(n) \quad\Rightarrow\quad d = \Omega(\log n)
\end{equation}
must hold.
Hence, the logarithmic depth lower bound is fundamental and unavoidable.

\begin{algorithm}[H]
\caption{Measurement-based randomized fan-out staircases}
\label{alg:staircase}
\begin{algorithmic}[1]
\State \textbf{Input:} System qubits $\{Q_i\}_{i=1}^{n}$, auxiliary qubits $\{\widetilde{Q}_j\}_{j=1}^{n-1}$, constants $\mathfrak{r}_1, \mathfrak{r}_2$, and number of paths $\mathfrak{D}$.
\For{$d=1,\dots,\mathfrak{D}$} \Comment{Iterate over $\mathfrak{D}$ directed paths}
\State Select a random directed qubit path $\{Q_{\ell_i}\}_{i=1}^{n}$, $\{\widetilde{Q}_{\ell_j}\}_{j=1}^{n-1}$.
\For{$r=1,\dots,\mathfrak{r}_1$}
\State Apply $\mathrm{CX}(Q_{\ell_i},\widetilde{Q}_{\ell_{i}})$ and $\mathrm{CX}(\widetilde{Q}_{\ell_i},Q_{\ell_{i+1}})$ sequentially along the path.
\EndFor 
\For{$r=1,\dots,\mathfrak{r}_2$}
\For{each system qubit $Q_{\ell_i}$}
\State Randomly select an auxiliary qubit $\widetilde{Q}_{\ell_j}$, $0 \leq j-i <n/\mathfrak{r}_2$, without replacement; apply $\mathrm{CX}(Q_{\ell_i},\widetilde{Q}_{\ell_j})$.
\EndFor
\For{each auxiliary qubit $\widetilde{Q}_{\ell_j}$ acted on}
\State Randomly select system qubit $Q_{\ell_k}$, $\mathfrak{r}_1 < k-j < n/\mathfrak{r}_2$, without replacement; apply $\mathrm{CX}(\widetilde{Q}_{\ell_j},Q_{\ell_k})$.
\EndFor
\EndFor \Comment{Forward fan-out ladder}
\State Measure auxiliary qubits in $X$-basis; store
 outcomes.
\State Reverse indexing order:
$Q_{\ell_i}\rightarrow Q_{\ell_{n+1-i}}, \widetilde{Q}_{\ell_j}\rightarrow \widetilde{Q}_{\ell_{n-j}}$.
\State Repeat steps 4--15 using reversed indices.
\EndFor \Comment{Backward fan-out ladder}
\State \textbf{Feed-forward:} Compute transfer matrices $\{\mathcal{T}^{(i)}\}_{i=1}^{2\mathfrak{D}}$ and apply corresponding Pauli-$Z$ corrections to system qubits.
\State \textbf{Output:} A $2\mathfrak{D}$-layer fan-out staircase circuit comprising $\mathcal{O}(\mathfrak{D}n)$ fan-out gates, each targeting $\mathcal{O}(\mathfrak{r}_1+4^{\mathfrak{r}_2})$ qubits.
\end{algorithmic}
\end{algorithm}

In the special scenario where each fan-out gate in the fan-out ladder targets a fixed, contiguous set of qubits—specifically, for each control qubit \(Q_{\ell_i}\), the target set is \(T_{\ell_i} = \{Q_{\ell_{i+1}},Q_{\ell_{i+2}},\dots,Q_{\ell_{i+k}}\}\) for a finite integer \(k\)—the corresponding \(\mathrm{CX}\) circuit can be represented as a matrix \(\mathbf{M} \in \operatorname{GL}(n, 2)\) with entries equal to 1 only on the diagonal and within a band of width \(k\) around the diagonal. Under these conditions, the original \(\mathcal{O}(n)\)-depth fan-out ladder circuit can be efficiently compiled into an equivalent \(\mathcal{O}(\log n)\)-depth circuit without auxiliary qubits, as follows:

Initially, apply sequential \(\mathrm{CX}\) gates between neighboring qubits: explicitly, for \(i=1,2,\dots,n-1\), apply \(\mathrm{CX}(Q_{\ell_i}, Q_{\ell_{i+1}})\). Subsequently, apply \(\mathrm{CX}\) gates between qubits separated by distance  \(k\) in reverse order: specifically, for \(i=n-k, n-k-1,\dots,1\), apply \(\mathrm{CX}(Q_{\ell_i}, Q_{\ell_{i+k}})\). This procedure yields an equivalent circuit to the original contiguous fan-out ladder.
This compiled circuit consists of $(k+1)$ sequential \(\mathrm{CX}\) ladders, each compilable into $\mathcal{O}(\log n)$ depth without auxiliary qubits, as demonstrated in Ref.~\cite{Remaud2025Ancilla}. Consequently, the complete fan-out ladder can be realized as a \(\mathrm{CX}\) circuit of depth $\mathcal{O}(k\log n)$. \hfill$\blacksquare$

\noindent Ref.~\cite{Buhrman2024State} shows that other FS realized by constant-depth dynamic circuits also compile to the logarithmic depth bound on all-to-all hardware, albeit with more complex constructions.

Each of the $2\mathfrak{D}$ fan-out ladders involves mid-circuit measurements yielding binary outcomes $\{m_{j}\}_{j=1}^{n-1}$. A measurement outcome $m_{j}=1$ is equivalent to the scenario of applying a Pauli-$Z$ operator to the $\widetilde{Q}_j$ before its Hadamard measurement, effectively converting the outcome to 0. Using propagation relations \(\mathrm{CX}(Q_i,Q_j) Z_j = Z_iZ_j\mathrm{CX}(Q_i,Q_j)\), these hypothetical $Z$ operations translate into feed-forward corrections exclusively on system qubits. The necessary corrections after measurements can be succinctly expressed using an $n\times(n-1)$ binary transfer matrix $\mathcal{T}$ as
\begin{equation}
\bigotimes_i Z^{(\mathcal{T}\mathbf{m})_i},\ \text{with} \ \ (\mathcal{T}\mathbf{m})_i \coloneqq \sum_j \mathcal{T}_{i,j} m_j \bmod 2.
\end{equation}
Computing $\mathcal{T}$ across all $2\mathfrak{D}$ rounds has complexity $\mathcal{O}(\mathfrak{D}(\mathfrak{r}_1 + \mathfrak{r}_2)n^2)$. Rather than immediate corrections, we propagate all corrections to circuit end using relations 
\begin{equation}
\begin{aligned}
&\textsc{FanOut}(Q_i\to T_i) Z_{Q_j}\\ &=
\begin{cases}
Z_{Q_i}Z_{Q_j}\textsc{FanOut}(Q_i\to T_i), & Q_j\in T_i, \\[4pt]
Z_{Q_j}\textsc{FanOut}(Q_i\to T_i), & Q_j\notin T_i,
\end{cases}
\end{aligned}
\end{equation}
to propagate all corrections to the end of the staircase circuit. An illustrative example follows, with further details and examples provided in the SM.

\begin{example}[Transfer matrix for a CX ladder~\cite{Baumer2024Measurement}]
Consider $n$ system qubits $\{Q_i\}$ and $n-1$ auxiliary qubits $\{\widetilde Q_i\}$ arranged linearly. 
Apply two nearest-neighbor layers: first $\mathrm{CX}(Q_i,\widetilde Q_i)$ for $i=1,\dots,n-1$, then $\mathrm{CX}(\widetilde Q_i,Q_{i+1})$ for $i=1,\dots,n-1$. 
Measure all $\widetilde Q_i$ in the $X$ basis with outcomes $m_i\in\{0,1\}$. 
Using $\mathrm{CX}(c,t)\,Z_t=Z_cZ_t\,\mathrm{CX}(c,t)$ and $Z |0\rangle = |0\rangle$, a putative $Z_{\widetilde Q_i}$ back-propagates to $Z_{Q_1}Z_{Q_2}\cdots Z_{Q_i}$, so the system-side Pauli-frame bits are
\begin{equation}
z_j = \bigoplus_{k=j}^{n-1} m_k, \quad \text{for } j=1, \dots, n-1
\end{equation}
and \(z_n=0\). Equivalently, $\mathbf z=\mathcal T\mathbf m\ (\mathrm{mod}\ 2)$ with $\mathcal T\in\{0,1\}^{n\times(n-1)}$ being the upper-triangular matrix (suffix-parity). 
For $n=4$,
\begin{equation}
\mathcal T=\begin{pmatrix}
1 & 1 & 1\\
0 & 1 & 1\\
0 & 0 & 1\\
0 & 0 & 0
\end{pmatrix}.
\end{equation}
\end{example}

\begin{figure}[tbh]
  \centering
  \includegraphics[width=7.7cm]{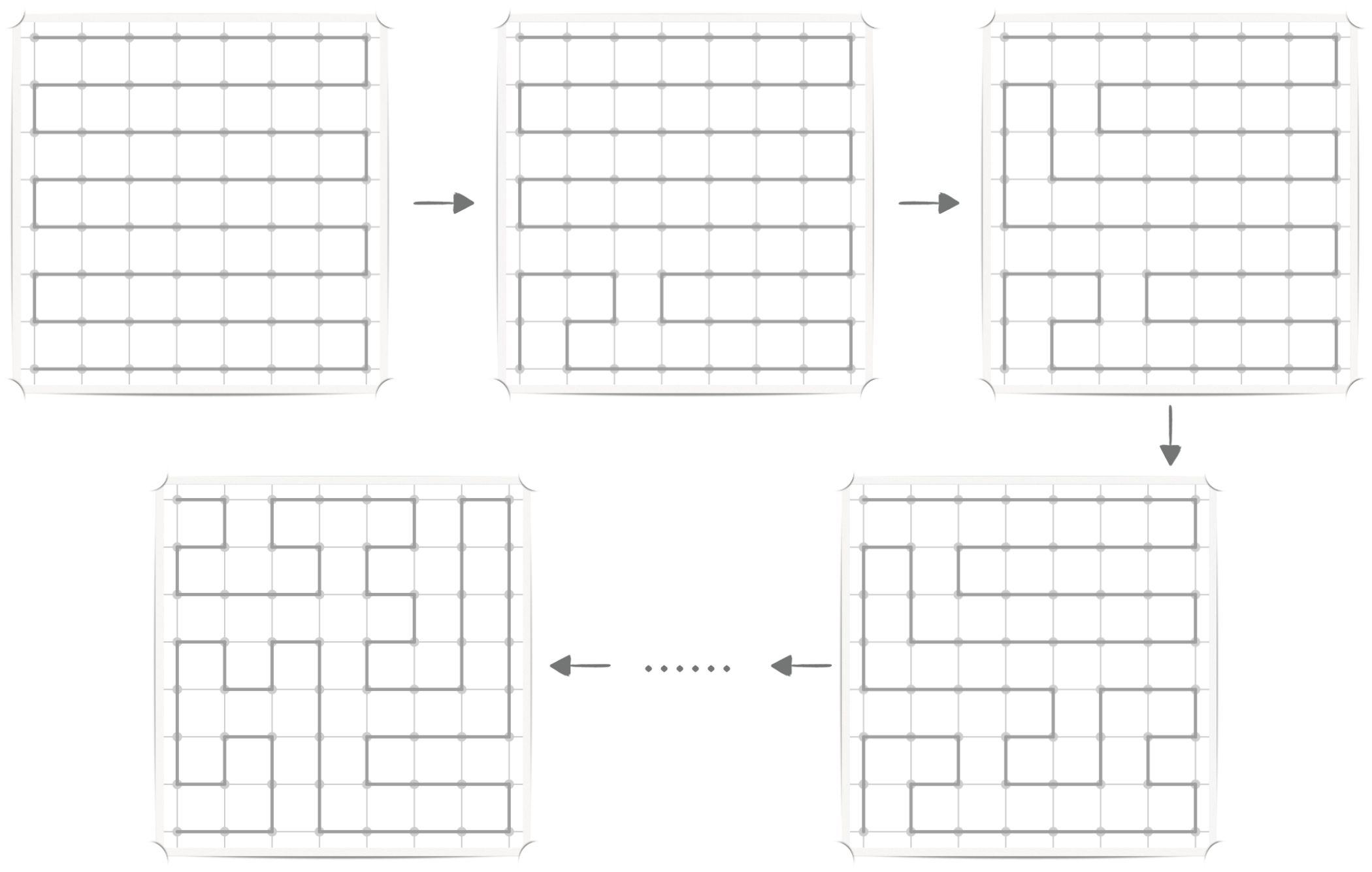}
  \caption{An example of the split-and-mend procedure for constructing a random directed Hamiltonian path on a 2D grid.}
  \label{fig:hamiltonian_path}
\end{figure}

On a 2D grid, the exponential number of Hamiltonian paths allows us to generate a distinct, effectively deep, constant-depth IQP circuit for each sampling round. We generate these paths as detailed in Protocol~\ref{alg:hamiltonian_path} (e.g., Fig.~\ref{fig:hamiltonian_path}), using 2000 rerouting iterations in our simulations.

\begin{algorithm}[H]
\caption{Random directed Hamiltonian path on a 2D grid}
\label{alg:hamiltonian_path}
\begin{algorithmic}[1]
\State \textbf{Input:} Grid dimensions $(\mathsf{width}, \mathsf{height})$, start qubit $(x_0, y_0)$, and number of re-routing iterations $\mathsf{count}$.
\State \textbf{Initialization:}
\begin{itemize}
  \item For each qubit $(x,y)$, assign an initial zig-zag direction:
        \begin{itemize}
            \item If $y$ is even, set $\mathsf{dir}(x,y)\leftarrow\mathsf{WEST}$, except for $x=0$, set $\mathsf{dir}(0,y)\leftarrow\mathsf{NORTH}$.
            \item If $y$ is odd, set $\mathsf{dir}(x,y)\leftarrow\mathsf{EAST}$, except for $x=\mathsf{width}-1$, set $\mathsf{dir}(x,y)\leftarrow\mathsf{NORTH}$.
        \end{itemize}
  \item Set $\mathsf{dir}(x_0,y_0)\leftarrow\mathsf{NOWHERE}$, ensuring the start qubit has no incoming edge.
\end{itemize}
\For{$i = 1,\dots,\mathsf{count}$} \Comment{Iterative reconfiguration}
  \State \textbf{Split:} Select a $2\times2$ sub-square whose two diagonals carry opposite directions.  
        Let one diagonal be $(p,q)$; follow stored directions from $p$ until $q$ is reached,  
        marking the square as a loop.  If this fails, switch to the other diagonal $(p',q')$ and retry.
  \State \textbf{Modify$_1$:} Flip $(p,q)$ from vertical 
        ($\{\mathsf{NORTH},\mathsf{SOUTH}\}$) to horizontal 
        ($\{\mathsf{EAST},\mathsf{WEST}\}$), or vice versa.
  \State \textbf{Mend:} Choose a diagonal pair $(u,v)$ with exactly one endpoint inside the loop;  
        if none exists, set $(u,v)\leftarrow(p,q)$.
  \State \textbf{Modify$_2$:} Apply the same flip to $(u,v)$ to merge the loop into the global path.
\EndFor
\State \textbf{Extract path:}
\begin{itemize}
  \item Compute the in-degree of each qubit.
  \item Select any qubit with in-degree $0$ and follow its edges to obtain a single chain covering 
  all qubits.
\end{itemize}
\State \textbf{Output:} A directed Hamiltonian path that visits each qubit 
exactly once.
\end{algorithmic}
\end{algorithm}

\clearpage
\twocolumngrid
\appendix
\begin{center}
	\textbf{\large Supplementary Material for ``Measurement-driven quantum advantages in shallow circuits"}
\end{center}
\setcounter{equation}{0}
\setcounter{figure}{0}
\setcounter{table}{0}
\setcounter{page}{1}
\makeatletter
\renewcommand{\thetable}{S\arabic{table}}
\renewcommand{\thefigure}{S\arabic{figure}}
\renewcommand{\theequation}{S\arabic{equation}}

\section{Appendix A: Fan-out ladders and staircases}\label{Appendix:Fan-outLadders}
\begin{figure*}[tbh]
	\centering
    \includegraphics[width=18cm]{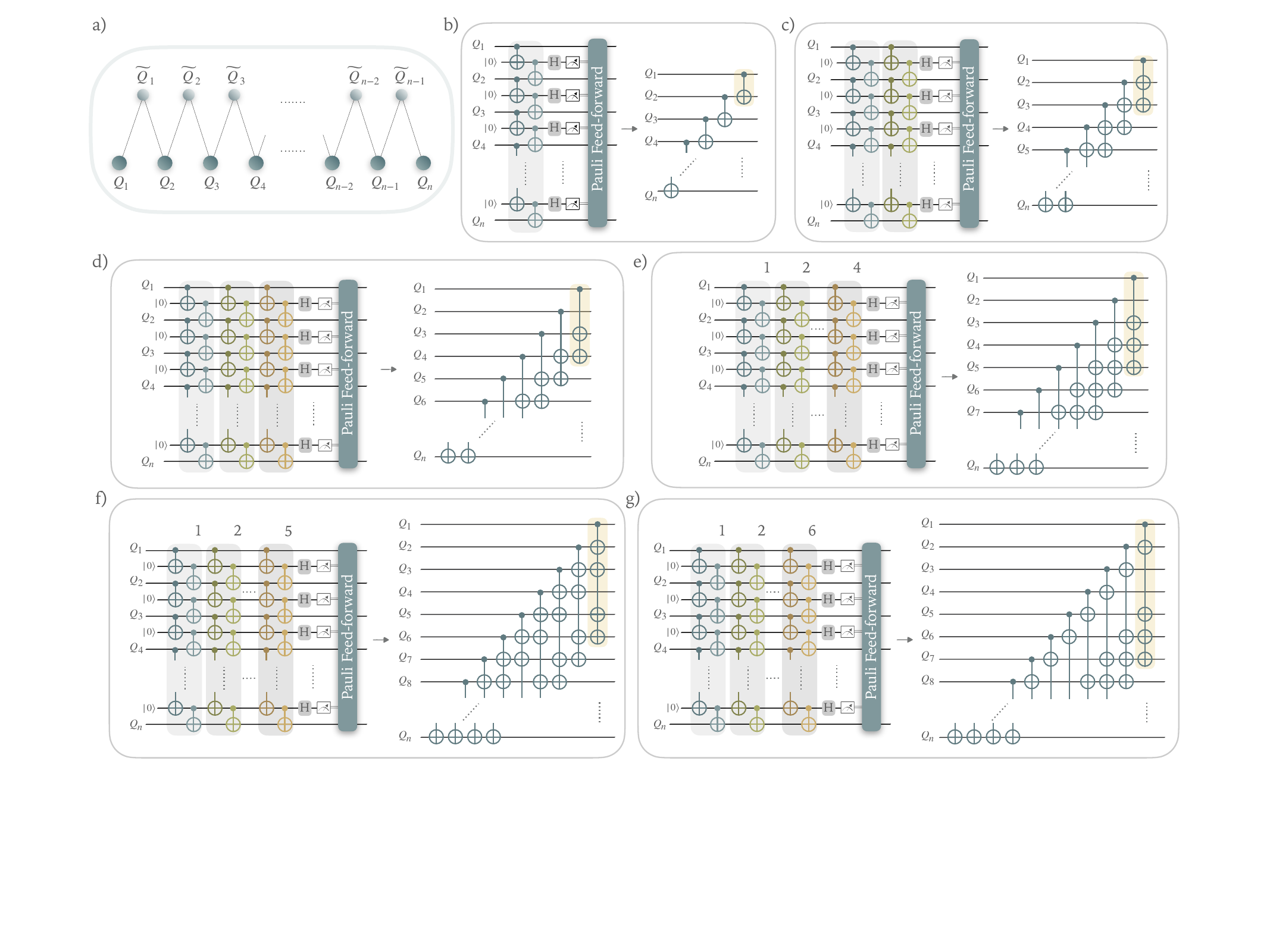}
	\caption{(a) Schematic of a one-dimensional topology where system qubits and auxiliary qubits alternate. (b) A \(\mathrm{CX}\) ladder, denoted $\mathsf{FL}_{1}$, implemented in a 1D topology with one layer. (c–g) Fan-out ladders denoted $\mathsf{FL}^{(\uparrow)}_{(1,2)}$,$\mathsf{FL}^{(\uparrow)}_{(2,3)}$,$\mathsf{FL}^{(\uparrow)}_{(2,3,4)}$,$\mathsf{FL}^{(\uparrow)}_{(1,2,4,5)}$, and $\mathsf{FL}^{(\uparrow)}_{(1,4,5,6)}$ are achieved through 2, 3, 4, 5, and 6 repetitions of nearest-neighbor \(\mathrm{CX}\) layers, respectively. In these 1D implementations, each successive fan-out operation does not necessarily target a contiguous block of qubits—some intermediate connections are partially canceled. Nevertheless, for a fixed number of layers and away from circuit boundaries, the overall structure retains translational symmetry.}
	\label{fig:Fanout-ladder-1D}
\end{figure*}

\begin{figure*}[tbh]
	\centering
    \includegraphics[width=18cm]{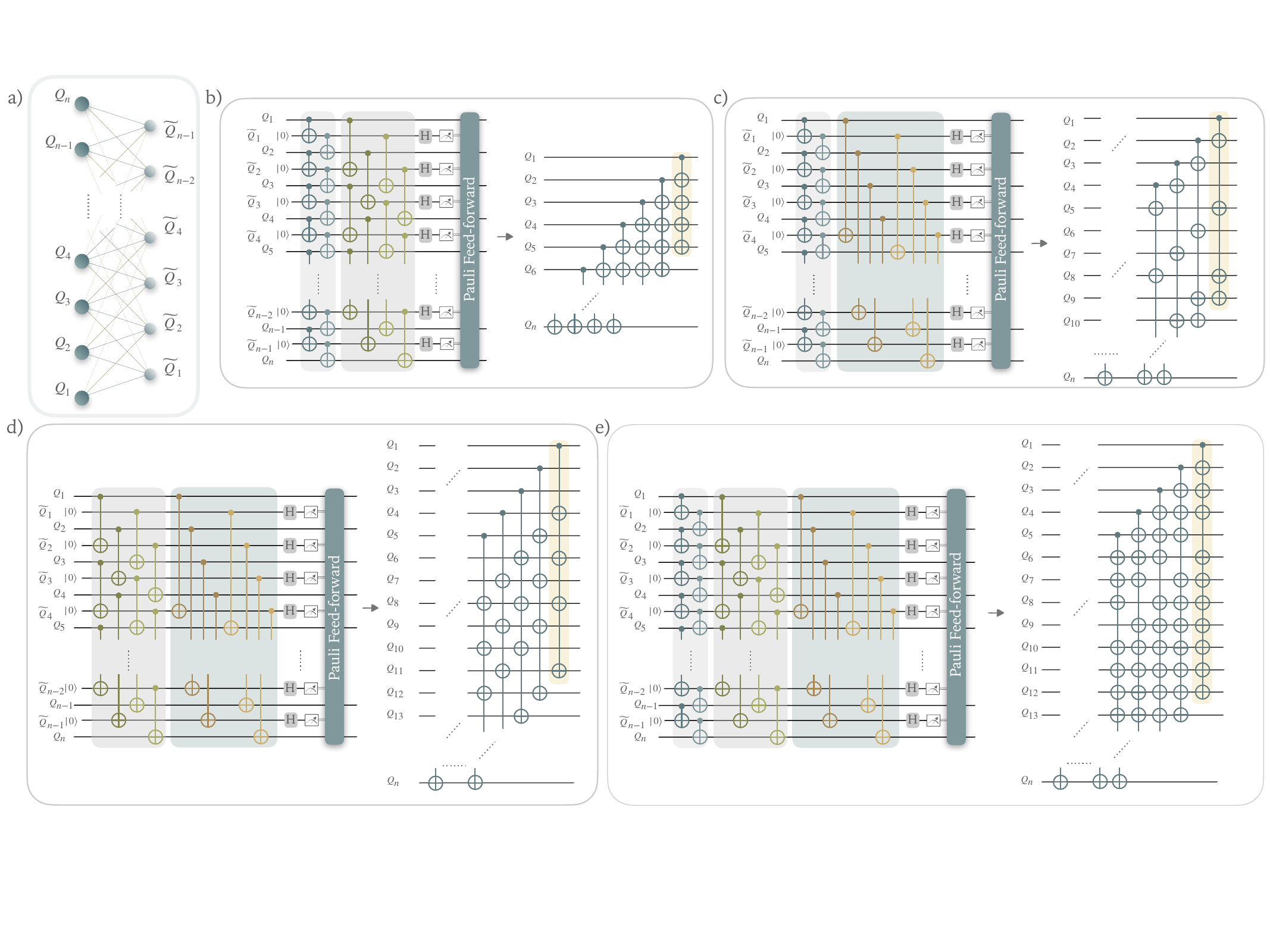}
	\caption{(a) A bipartite graph topology in which each system qubit interacts with multiple nearest‐neighbor auxiliary qubits. (b) A dynamic circuit realizing the fan‐out ladder $\mathsf{FL}^{(\uparrow)}_{(1,2,3,4)}$, where each system qubit $Q_i$ is connected to auxiliary qubits $\widetilde{Q}_{i - 2}$, $\widetilde{Q}_{i - 1}$, $\widetilde{Q}_{i}$, and $\widetilde{Q}_{i + 1}$ (when available). (c) A dynamic circuit realizing the fan‐out ladder $\mathsf{FL}^{(\uparrow)}_{(1,4,7,8)}$, where each system qubit $Q_i$ is connected to auxiliary qubits $\widetilde{Q}_{i - 4}$, $\widetilde{Q}_{i - 1}$, $\widetilde{Q}_{i}$, and $\widetilde{Q}_{i + 3}$ (when available). (d) A dynamic circuit realizing the fan‐out ladder $\mathsf{FL}^{(\uparrow)}_{(3,5,7,10)}$, where each system qubit $Q_i$ is connected to auxiliary qubits $\widetilde{Q}_{i - 4}$, $\widetilde{Q}_{i - 2}$, $\widetilde{Q}_{i + 1}$, and $\widetilde{Q}_{i + 3}$ (when available). (e) A dynamic circuit realizing the fan‐out ladder $\mathsf{FL}^{(\uparrow)}_{(1,2,3,5,\ldots,12)}$, where each system qubit $Q_i$ is connected to auxiliary qubits $\widetilde{Q}_{i - 4}$, $\widetilde{Q}_{i - 2}$, $\widetilde{Q}_{i - 1}$, $\widetilde{Q}_{i}$, $\widetilde{Q}_{i + 1}$, and $\widetilde{Q}_{i + 3}$ (when available). In all cases, the structure of each fan‐out gate is determined by the propagation of information in the original circuit, and the overall configuration retains translational symmetry away from circuit boundaries.}
	\label{fig:Fanout-ladder-2}
\end{figure*}

\begin{figure*}[tbh]
	\centering
\includegraphics[width=18cm]{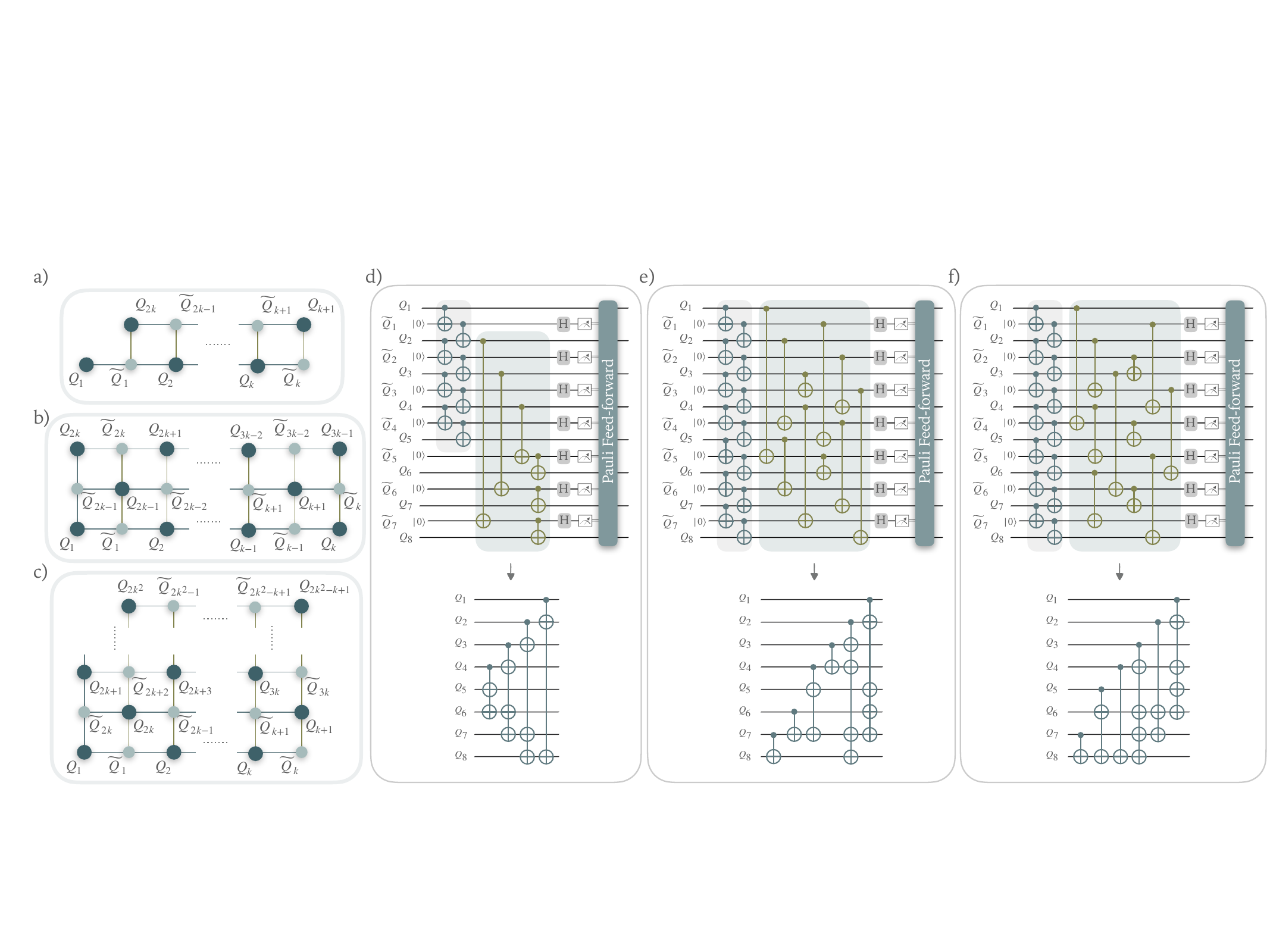}
	\caption{(a) A two-layer grid topology with alternating system and auxiliary qubits, containing $2k$ system qubits and $2k-1$ auxiliary qubits for an integer $k$. (b) A three-layer grid topology with alternating system and auxiliary qubits, 
    containing $3k-1$ system qubits and $3k-1$ auxiliary qubits for an integer $k$. (c) A $2k \times 2k$ grid topology (with one missing corner) containing $2k^2$ system qubits and $2k^2-1$ auxiliary qubits. (d) An example 15-qubit dynamic circuit based on the grid topology in (a) that realizes a non-translationally invariant (irregular) fan-out ladder. (e) An example 15-qubit dynamic circuit based on the grid topology in (b) that realizes an irregular fan-out ladder. (f) An example 15-qubit dynamic circuit based on the grid topology in (c) that realizes an irregular fan-out ladder.}
	\label{fig:Fanout-ladder-3}
\end{figure*}

Fan-out staircases are constructed by concatenating multiple fan-out ladders, with adjacent ladders arranged in opposite orders. In this appendix, we describe in detail how to construct fan-out ladders that implement various fan-out gates under different connectivity topologies. We also explain how these fan-out ladders can be combined to form a fan-out staircase that incorporates multiple rounds of mid-circuit measurement while requiring only a single round of feed-forward.
Figs.~\ref{fig:Fanout-ladder-1D}, \ref{fig:Fanout-ladder-2}, and \ref{fig:Fanout-ladder-3} illustrate some fan-out ladder examples in \emph{one-dimensional} (1D) connectivity, bipartite graph connectivity, and \emph{two-dimensional} (2D) grid topologies, respectively.

\subsubsection{One-dimensional topology}
Consider first the 1D connectivity case, where system qubits and auxiliary qubits alternate, as shown in Fig.~\ref{fig:Fanout-ladder-1D}(a). It is well known that a single layer of nearest-neighbor \(\mathrm{CX}\) operations, combined with mid-circuit measurements, can implement a \(\mathrm{CX}\) ladder [Fig.~\ref{fig:Fanout-ladder-1D}(b)]. (Apparently, the inverse of this ladder can be used to prepare GHZ states.) By repeating these operations multiple times, one obtains fan-out ladders that—despite partial cancellation of some intermediate connections—retain overall translational symmetry away from circuit boundaries. The detailed structure of the required feed-forward operations is determined by the circuit connectivity. In what follows, we rigorously derive the equivalent transformation implemented by the \(\mathrm{CX}\)/fan-out ladder and characterize the corresponding feed-forward corrections based on the measurement outcomes for one and two repetitions of the \(\mathrm{CX}\) layers.

\paragraph{Single repetition of the \(\mathrm{CX}\) layer.}
Let the system qubits be denoted by
\begin{equation}
Q_1,\, Q_2,\, \dots,\, Q_n,
\end{equation}
and the auxiliary qubits by
\begin{equation}
\widetilde{Q}_1,\, \widetilde{Q}_2,\, \dots,\, \widetilde{Q}_{n-1},
\end{equation}
with each auxiliary qubit initially prepared in the state vector $|0\rangle$. We denote the controlled-NOT (CX) gate with control qubit $Q_i$ and target qubit $\widetilde{Q}_j$ as \(\mathrm{CX}(Q_i,\widetilde{Q}_j)\), and the reverse operation,  where \(\widetilde{Q}_j\) is the control and \(Q_i\) is the target, as \(\mathrm{CX}(\widetilde{Q}_j, Q_i)\).

The overall unitary corresponding to a single repetition of the \(\mathrm{CX}\) layer is given by
\begin{equation}
U^{(1)} =  \left(\prod_{i=1}^{n-1} \mathrm{CX}(\widetilde{Q}_i, Q_{i+1})\right)\cdot \left(\prod_{i=1}^{n-1} \mathrm{CX}(Q_i,\widetilde{Q}_i)\right).
\end{equation}

We now show that if, for a given auxiliary qubit \(\widetilde{Q}_i\) (initialized in \(|0\rangle\)), we sequentially apply \(\mathrm{CX}(Q_i,\widetilde{Q}_i)\) and \(\mathrm{CX}(\widetilde{Q}_i, Q_{i+1})\), then subsequently applying a Hadamard gate on \(\widetilde{Q}_i\) and measuring \(\widetilde{Q}_i\) in the computational basis with outcome \(|0\rangle\) yields an effective operation on the system qubits equivalent to a \(\mathrm{CX}\) gate from \(Q_i\) to \(Q_{i+1}\). For clarity, we now detail the state evolution.

Assume that the initial state vector of the three qubits involved is
\begin{equation}
|\Psi_0\rangle = \Bigl(\alpha\,|0\rangle_{Q_i} + \beta\,|1\rangle_{Q_i}\Bigr) \otimes |0\rangle_{\widetilde{Q}_i} \otimes \Bigl(\gamma\,|0\rangle_{Q_{i+1}} + \delta\,|1\rangle_{Q_{i+1}}\Bigr),
\end{equation}
where \(\alpha, \beta, \gamma, \delta \in \mathbb{C}\) satisfy the normalization conditions. 

The application of the first gate, \(\mathrm{CX}(Q_i,\widetilde{Q}_i)\), transforms the state vector into
\begin{equation}
\begin{aligned}
|\Psi_1\rangle ={} & \alpha\,|0\rangle_{Q_i}\,|0\rangle_{\widetilde{Q}_i}\,\Bigl(\gamma\,|0\rangle_{Q_{i+1}} + \delta\,|1\rangle_{Q_{i+1}}\Bigr)\\
&+ \beta\,|1\rangle_{Q_i}\,|1\rangle_{\widetilde{Q}_i}\,\Bigl(\gamma\,|0\rangle_{Q_{i+1}} + \delta\,|1\rangle_{Q_{i+1}}\Bigr).
\end{aligned}
\end{equation}
Subsequently, the application of the second gate, \(\mathrm{CX}(\widetilde{Q}_i, Q_{i+1})\), results in the state vector
\begin{equation}
\begin{aligned}
|\Psi_2\rangle ={} & \alpha\,|0\rangle_{Q_i}\,|0\rangle_{\widetilde{Q}_i}\,\Bigl(\gamma\,|0\rangle_{Q_{i+1}} + \delta\,|1\rangle_{Q_{i+1}}\Bigr) \\
&+ \beta\,|1\rangle_{Q_i}\,|1\rangle_{\widetilde{Q}_i}\,\Bigl(\gamma\,|1\rangle_{Q_{i+1}} + \delta\,|0\rangle_{Q_{i+1}}\Bigr).
\end{aligned}
\end{equation}
We now apply the Hadamard gate on \(\widetilde{Q}_i\) and project onto the subspace where the auxiliary qubit is measured in the computational basis with outcome \(|0\rangle\). Discarding the normalization factor, the resulting (post-measurement) state is

\begin{equation}
\begin{aligned}
|\Psi_f\rangle \propto {} & \alpha\,|0\rangle_{Q_i}\,\Bigl(\gamma\,|0\rangle_{Q_{i+1}} + \delta\,|1\rangle_{Q_{i+1}}\Bigr) \\
&+ \beta\,|1\rangle_{Q_i}\,\Bigl(\gamma\,|1\rangle_{Q_{i+1}} + \delta\,|0\rangle_{Q_{i+1}}\Bigr).
\end{aligned}
\end{equation}

This final state is equivalent to the state obtained by directly applying a \(\mathrm{CX}\) gate with \(Q_i\) as control and \(Q_{i+1}\) as target to the initial two-qubit state. If all auxiliary measurement outcomes are \(|0\rangle\), the overall effect is a \(\mathrm{CX}\) ladder, as illustrated in Fig.~\ref{fig:Fanout-ladder-1D}(b).

Now we consider the more general case in which, after applying \(U^{(1)}\) and the Hadamard, the measurement outcome on a particular auxiliary qubit \(\widetilde{Q}_j\) is \(|1\rangle\). This outcome is equivalent to having first applied a $Z$ gate to \(\widetilde{Q}_j\) prior to the Hadamard transformation, followed by measurement, which then yields the outcome \(|0\rangle\). Using the propagation relations for Pauli-$Z$ operators under conjugation by \(\mathrm{CX}\) gates
gives
\begin{equation}\label{eq:prop_relation1}
\begin{aligned}
    &Z(\widetilde{Q}_j)\left(\prod_i^{n-1} \mathrm{CX}(Q_i, \widetilde{Q}_i)\right) \\= &\left(\prod_i^{n-1} \mathrm{CX}(Q_i, \widetilde{Q}_i)\right)Z(Q_j)Z(\widetilde{Q}_j),
\end{aligned}
\end{equation}
and
\begin{equation}\label{eq:prop_relation2}
\begin{aligned}
    &Z(\widetilde{Q}_{j-1})Z(Q_j)\left(\prod_{i=1}^{n-1} \mathrm{CX}(\widetilde{Q}_i, Q_{i+1})\right) \\= &\left(\prod_{i=1}^{n-1} \mathrm{CX}(\widetilde{Q}_i, Q_{i+1})\right)Z(Q_j),
\end{aligned}
\end{equation}
one obtains, by successive application,
\begin{equation}
\begin{aligned}
    Z(\widetilde{Q}_j)U^{(1)} =& U^{(1)}Z(Q_j)Z(\widetilde{Q}_j)\\=&Z(\widetilde{Q}_{j-1})Z(Q_j)U^{(1)}Z(\widetilde{Q}_j)\\=&Z(\widetilde{Q}_{j-2})Z(Q_{j-1})Z(Q_j)U^{(1)}Z(\widetilde{Q}_{j-1})Z(\widetilde{Q}_j)\\=&\cdots\\=&\left(\prod_{i=1}^{j}Z(Q_{i}) \right)U^{(1)}\left(\prod_{i=1}^{j}Z(\widetilde{Q}_{i}) \right).
\end{aligned}
\end{equation}
Since the initial state vectors of the auxiliary qubits are \(|0\rangle\), the application of \(\left(\prod_{i=1}^{j}Z(\widetilde{Q}_{i}) \right)\) does not alter their state. Consequently, obtaining a measurement outcome \(m_j=1\) on \(\widetilde{Q}_j\) is equivalent to the case \(m_j=0\) up to the application of corrective $Z$ operations on the system qubits \(Q_1, Q_2, \dots, Q_j\). 

More generally, we can encapsulate the required feed-forward correction in terms of a binary transfer matrix \(\mathcal{T}\) that is determined by the circuit structure. In particular, for each system qubit \(Q_i\), the corrective \(Z\) operation can be written as
\begin{equation}
Z^{f_i} \quad \text{with} \quad f_i = \bigoplus_j  \mathcal{T}_{i,j}\, m_j,
\end{equation}
where \(m_j\) is the measurement outcome (0 or 1) for the auxiliary qubit \(\widetilde{Q}_j\), and the binary matrix \( \mathcal{T}\) encodes the propagation of the \(Z\) operators through the circuit. In other words, the feed-forward correction on the system qubits is completely determined by the parity of the measured outcomes on the corresponding auxiliary qubits, as dictated by the transfer matrix $\mathcal{T}$. For the one-layer case, the transfer matrix $\mathcal{T}$ is given by:
\begin{equation}
     \mathcal{T}_{i , j}= \begin{cases}1, & \text { if } i\le j, \\ 0, & \text { if } i > j,\end{cases}
\end{equation}
where \(i\) and \(j\) denote the indices of the system qubits and auxiliary qubits, respectively.

\paragraph{Two repetitions of the \(\mathrm{CX}\) layer.}
We now extend the analysis to the case of two repetitions of the nearest-neighbor \(\mathrm{CX}\) layers. The overall unitary for two repetitions is given by
\begin{equation}
    U^{(2)} = U^{(1)}\cdot U^{(1)}.
\end{equation}
To illustrate the construction, we focus on a block comprising five qubits: system qubits $Q_i$, $Q_{i+1}$, $Q_{i+2}$, and auxiliary qubits $\widetilde{Q}_i$, $\widetilde{Q}_{i+1}$. The sequence of \(\mathrm{CX}\) gates acting on these qubits is
\begin{equation}
\begin{aligned}
&\mathrm{CX}(Q_i, \widetilde{Q}_i), \quad
\mathrm{CX}(Q_{i+1}, \widetilde{Q}_{i+1}),\\
&\mathrm{CX}(\widetilde{Q}_i, Q_{i+1}), \quad
\mathrm{CX}(\widetilde{Q}_{i+1}, Q_{i+2}),\\
&\mathrm{CX}(Q_{i+1}, \widetilde{Q}_{i+1}), \quad
\mathrm{CX}(\widetilde{Q}_{i+1}, Q_{i+2}).
\end{aligned}
\end{equation}

After these \(\mathrm{CX}\) operations, Hadamard gates are applied to both  $\widetilde{Q}_i$ and $\widetilde{Q}_{i+1}$  followed by measurement in the computational basis.

For clarity, let the initial state vector of these five qubits be
\begin{equation}
\begin{aligned}
|\Psi_0\rangle =\,& |\psi\rangle_{Q_i} \otimes |0\rangle_{\widetilde{Q}_i} \otimes |\phi\rangle_{Q_{i+1}} \otimes |0\rangle_{\widetilde{Q}_{i+1}} \otimes |\chi\rangle_{Q_{i+2}},
\end{aligned}
\end{equation}
with 
\(|\psi\rangle_{Q_i} = \alpha\,|0\rangle + \beta\,|1\rangle\),
\(|\phi\rangle_{Q_{i+1}} = \gamma |0\rangle + \delta \,|1\rangle\),
\(|\chi\rangle_{Q_{i+2}} = \gamma' |0\rangle + \delta' \,|1\rangle\). 
where $\alpha, \beta, \gamma, \delta, \gamma^{\prime}, \delta^{\prime} \in \mathbb{C}$ and the states are normalized.

Assuming that the Hadamard measurements on both auxiliary systems yield \(|0\rangle\) (i.e., \(m_i=0\) and \(m_{i+1}=0\)), the post-selected state vector becomes
\begin{equation}
\begin{aligned}
|\Psi_f\rangle \propto {} & \alpha\gamma\,|0\rangle_{Q_i}\,|0\rangle_{Q_{i+1}}\,\Bigl(\gamma'\,|0\rangle_{Q_{i+2}} + \delta'\,|1\rangle_{Q_{i+2}}\Bigr) \\
&+ \alpha\delta\,|0\rangle_{Q_i}\,|1\rangle_{Q_{i+1}}\,\Bigl(\delta'\,|0\rangle_{Q_{i+2}} + \gamma'\,|1\rangle_{Q_{i+2}}\Bigr)\\
&+ \beta\delta\,|1\rangle_{Q_i}\,|0\rangle_{Q_{i+1}}\,\Bigl(\gamma'\,|0\rangle_{Q_{i+2}} + \delta'\,|1\rangle_{Q_{i+2}}\Bigr)\\
&+ \beta\gamma\,|1\rangle_{Q_i}\,|1\rangle_{Q_{i+1}}\,\Bigl(\delta'\,|0\rangle_{Q_{i+2}} + \gamma'\,|1\rangle_{Q_{i+2}}\Bigr).
\end{aligned}
\end{equation}
The net effect of these operations is equivalent to first applying a \(\mathrm{CX}\) gate from \(Q_{i+1}\) to \(Q_{i+2}\), followed by a fan-out gate with control \(Q_i\) and target qubits \(Q_{i+1}\) and \(Q_{i+2}\). By incorporating additional \(\mathrm{CX}\) layers over a larger set of qubits, we obtain a deeper fan-out ladder. In the two-layer case, if all measurement outcomes are $|0\rangle$, the resulting fan-out ladder—denoted \(\mathsf{FL}_{1,2}\)—is shown in Fig.~\ref{fig:Fanout-ladder-1D}(c).

A similar analysis applies to the feed-forward operations when the measurement outcome on a particular auxiliary qubit \(\widetilde{Q}_j\) is \(|1\rangle\). As before, this outcome is equivalent to having first applied a $Z$ gate to \(\widetilde{Q}_j\) before the Hadamard transformation, thereby converting the outcome to an effective \(|0\rangle\) measurement while introducing extra $Z$ operators on the system qubits.

Using the propagation relations for Pauli-$Z$ operators under conjugation by \(\mathrm{CX}\) gates, i.e., Eqs.~\eqref{eq:prop_relation1},~\eqref{eq:prop_relation2}, we have
\begin{equation}
\begin{aligned}
    Z(\widetilde{Q}_j)U^{(2)} =& U^{(1)}Z(Q_j)Z(\widetilde{Q}_j)U^{(1)}\\=& U^{(1)}Z(Q_j)U^{(1)}Z(Q_j)Z(\widetilde{Q}_j) \\=& U^{(1)}Z(\widetilde{Q}_{j-1})U^{(1)}Z(\widetilde{Q}_j)\\=& U^{(2)}Z(\widetilde{Q}_{j-1})Z(Q_{j})Z(\widetilde{Q}_j)\\=& Z(Q_{j-1})Z(Q_{j-2})Z(\widetilde{Q}_{j-3})U^{(2)}Z(\widetilde{Q}_{j-1})Z(\widetilde{Q}_j)\\=&\cdots\\=&\left(\prod_{i \in I_2(j)}Z(Q_{i}) \right) U^{(1)}\left(\prod_{i \in I_1(j)}Z(\widetilde{Q}_{i}) \right),
\end{aligned}
\end{equation}
where the index sets \(I_1(j)\) and \(I_2(j)\) capture the specific pattern of the propagation, defined by
\begin{equation}
I_1(j) \coloneqq
\left\{
\begin{aligned}
& j - 2k, \\
& j - 2k - 1
\end{aligned}
\;\middle|\;
k \in \mathbb{N}_0,  j - 2k \geq 1,  j - 2k - 1 \geq 1
\right\},
\end{equation}
and
\begin{equation}
I_2(j) \coloneqq 
\left\{
\begin{array}{l|l}
    j - 2k - 1,\, j - 2k - 2 
    & \begin{array}{l}
        k \in \mathbb{N}_0, \\
        j - 2k - 1 \geq 1, \\
        j - 2k - 2 \geq 1
      \end{array}
\end{array}
\right\}.
\end{equation}
Thus, obtaining a measurement outcome \(m_j=1\) on \(\widetilde{Q}_j\) is equivalent to the ideal outcome \(m_j=0\) up to the application of corrective $Z$ operations on the system qubits, specifically on those qubits whose indices belong to \(I_2(j)\). The transfer matrix here has entries
\begin{equation}
     \mathcal{T}_{i , j} = \begin{cases}1, & \text { if } i<j \text { and }(j-2) \bmod 3 \in\{1,2\}, \\ 0, & \text { otherwise.}\end{cases}
\end{equation}
This prescription determines the corrective $Z$ 
operation on each system qubit \(Q_i\) as a function of the parity of the measurement outcomes of a specific set of auxiliary systems.

\subsubsection{Generalization and optimization}
Deeper dynamic circuits—whether in 1D, on bipartite graphs, or on 2D grid topologies—can be analyzed using the formalism introduced earlier. For 1D topologies (see Fig.~\ref{fig:Fanout-ladder-1D}), fan-out ladders constructed with repeated \(\mathrm{CX}\) layers (e.g., $\mathsf{FL}_{(2,3)}$, $\mathsf{FL}_{(2,3,4)}$, $\mathsf{FL}_{(1,2,4,5)}$, and $\mathsf{FL}_{(1,4,5,6)}$, where subscript indices indicate the relative distance between the target qubits from the control qubits within each fan-out gate) generally exhibit translational symmetry away from circuit boundaries despite partial cancellation of intermediate connections. For bipartite connectivity graphs (Fig.~\ref{fig:Fanout-ladder-2}(a)), each system qubit interacts with multiple neighboring auxiliary qubits. Specifically, if system qubit $Q_i$ interacts with auxiliary qubits $\widetilde{Q}_{i-2}$, $\widetilde{Q}_{i-1}$, $\widetilde{Q}_i$, and $\widetilde{Q}_{i+1}$ (when available), the effective fan-out range increases more rapidly with additional layers. In 2D grid topologies, translational symmetry is typically lost. Fig.~\ref{fig:Fanout-ladder-3} shows illustrative examples of one-layer, two-layer, and square-grid topologies and their associated fan-out ladders. Our randomized dynamic circuits introduced earlier generalize these constructions, exhibiting an exponential increase in fan-out range with parameter $\mathfrak{r}_2$ when $\mathfrak{r}_2 \ll \log(n)$. The precise set of target qubits is determined dynamically based on information flow within the circuit.

Feed-forward corrections across all these structures are consistently products of identity and Pauli-$Z$ operators, analyzed using the transfer matrix formalism. Corrections from the first fan-out ladder propagate through subsequent ladders via commutation relationships between Pauli-$Z$ and \(\mathrm{CX}\) gates. After the final measurements, combined feed-forward corrections realize the complete fan-out staircase.

Now we discuss the synthesis and optimization of our randomized fan-out staircase circuits, specifically addressing whether our construction can be decomposed into shallower, more efficient circuits.

\begin{figure}[tb]
\centering
\includegraphics[width=8cm]{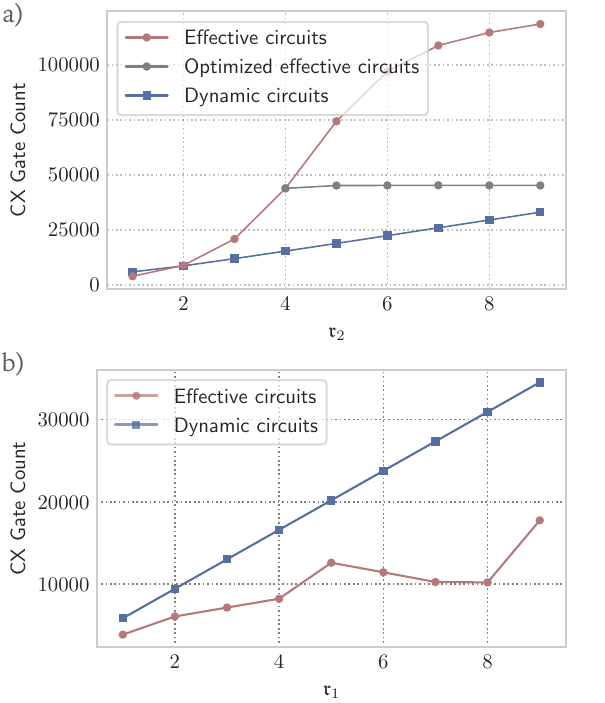}
\caption{Comparison of \(\mathrm{CX}\) (\(\mathrm{CNOT}\)) gate counts between original dynamic circuits, effective fan-out staircases after feed-forward corrections, and optimized effective circuits obtained via Gaussian elimination. (a) \(\mathrm{CX}\) gate counts for fixed $\mathfrak{r}_1=1$ and varying $\mathfrak{r}_2$. (b) \(\mathrm{CX}\) gate counts for fixed $\mathfrak{r}_2=1$ and varying $\mathfrak{r}_1$. Each data point represents the average over 50 randomly sampled instances, with error bars indicating standard deviations.}
\label{fig:CX count}
\end{figure}

Finally, we compare the number of \(\mathrm{CX}\) gates required by the original dynamic circuits, the corresponding effective fan-out staircases obtained via feed-forward corrections, and the optimized effective circuits derived through Gaussian elimination. We investigate whether these randomly generated fan-out staircases can be further optimized to reduce circuit complexity. Without loss of generality, numerical experiments are conducted with fixed parameters $n=300$ and $\mathfrak{D}=3$.

First, fixing $\mathfrak{r}_1=1$, we vary $\mathfrak{r}_2$, and sample 50 random circuit instances for each value. Results are shown in Fig.~\ref{fig:CX count}(a). Initially, for small $\mathfrak{r}_2$, the effective fan-out staircases require exponentially increasing numbers of \(\mathrm{CX}\) gates compared to the original dynamic circuits. At larger values of $\mathfrak{r}_2$, effective circuits become significantly denser. However, beyond a certain threshold ($\mathfrak{r}_2 > 4$), applying Gaussian elimination to convert the circuits into optimized effective forms leads to reductions in \(\mathrm{CX}\) count, revealing redundancy in the original staircase representation. Prior to this threshold, the Gaussian elimination method yields circuits with increased complexity, indicating no beneficial optimization.

Next, we fix $\mathfrak{r}_2=1$ and vary $\mathfrak{r}_1$ again sampling 50 random cases per parameter setting. As depicted in Fig.~\ref{fig:CX count}(b), the number of \(\mathrm{CX}\) gates in logical fan-out staircases increases significantly more slowly.


\section{Appendix B: Measurement-driven IQP sampling}\label{Appendix: HPS}
In the main text, we introduced \emph{measurement-driven IQP circuits}, constructed by interleaving randomized measurement-enabled fan-out staircase circuits with layers of single-qubit Pauli-$Z$ rotations. This appendix provides relevant proofs of theoretical claims, complemented by additional technical discussions and illustrative examples.

We first prove Proposition~1, 
which formally states that Hamiltonian phase states with random binary architectures achieve anticoncentration for a linear number of measurement and rotation steps.

\paragraph{Proof of Proposition 1.}
Consider the \(n\)-qubit Hamiltonian phase state vectors of the form
\begin{equation}
\lvert \psi_{\mathbf{A},\boldsymbol{\vartheta}} \rangle
=\exp\left(\mathrm{i}\sum_{i=1}^{s}\vartheta_i\bigotimes_{j=1}^{n}Z^{\mathbf{A}_{i,j}}\right) \lvert +\rangle^{\otimes n},
\end{equation}
where \(\mathbf{A}\in \{0,1\}^{s\times n}\) is a binary matrix whose ith row specifies the qubits on which the diagonal gate \(Z^{A_i}\) acts, and \(\boldsymbol{\vartheta} = (\vartheta_1,\dots,\vartheta_s)\) with each \(\vartheta_i\in[0,2\pi)\) drawn independently and uniformly. Here,
\begin{equation}
|+\rangle^{\otimes n}
\;\coloneqq\;
\frac{1}{\sqrt{2^n}}
\sum_{x\in\{0,1\}^n} |x\rangle.
\end{equation}

For a fixed architecture \(\mathbf{A}\), its two-copy moment operator is defined as
\begin{equation}
\rho^{(2)}_{\mathbf{A}}
\;\coloneqq\;
\mathbb{E}_{\boldsymbol{\vartheta}}
\Bigl[\lvert \psi_{\mathbf{A},\boldsymbol{\vartheta}}\rangle\langle\psi_{\mathbf{A},\boldsymbol{\vartheta}}\rvert^{\otimes2}\Bigr].
\end{equation}
Expressing \(\lvert \psi_{\mathbf{A},\boldsymbol{\vartheta}}\rangle\) in the computational basis, we obtain
\begin{equation}
\lvert \psi_{\mathbf{A},\boldsymbol{\vartheta}}\rangle
=\frac{1}{\sqrt{2^n}}\sum_{x\in\{0,1\}^n}
\exp \Bigl(\mathrm{i}\sum_{i=1}^s\vartheta_i\,(-1)^{A_i\cdot x}\Bigr)\,\lvert x\rangle,
\end{equation}
where \((-1)^{A_i\cdot x}\) captures parity \((A_i\cdot x\mod2)\). The second moment operator $|\psi_{\mathbf{A}, \boldsymbol{\vartheta}}\rangle\langle\psi_{\mathbf{A}, \boldsymbol{\vartheta}}|^{\otimes 2}$ then expands to
\begin{equation}
\begin{aligned}
\rho_{A,\boldsymbol{\vartheta}}^{\otimes2}
=
&\lvert \psi_{\mathbf{A},\boldsymbol{\vartheta}}\rangle\langle\psi_{\mathbf{A},\boldsymbol{\vartheta}}\rvert^{\otimes2}
\\=
&\frac{1}{2^{2n}}
\sum_{\substack{x,x^{\prime}\\y,y^{\prime}}}
\exp \Bigl(\mathrm{i}\sum_{i=1}^s \vartheta_i\,\bigl[(-1)^{A_i\cdot x}+(-1)^{A_i\cdot x^{\prime}}\\&-(-1)^{A_i\cdot y}-(-1)^{A_i\cdot y^{\prime}}\bigr]\Bigr)
\lvert x,x^{\prime}\rangle\langle y,y^{\prime}\rvert.
\end{aligned}
\end{equation}
Next, we average over the random phases \(\vartheta_i \in [0,2\pi)\). Recall that
\begin{equation}
    \int_0^{2\pi} e^{\mathrm{i}\alpha \vartheta}\,\frac{d\vartheta}{2\pi}
=
\delta_{\alpha,0},
\end{equation}
for integer \(\alpha\). Thus, \emph{any term in the sum} vanishes unless the exponent of each \(\vartheta_i\) is effectively zero, i.e.,
\begin{equation}
    (-1)^{A_i\cdot x} \;+\; (-1)^{A_i\cdot x^\prime}
\;=\;
(-1)^{A_i\cdot y} \;+\; (-1)^{A_i\cdot y^\prime}
\quad
\forall \,i.
\end{equation}
Equivalently, for each \(i\), the row \(A_i\) must fail to distinguish the pairs \((x,x^\prime)\) and \((y,y^\prime)\). 

The identity and the swap contributions certainly survive. Indeed, if \((x,y) = (x^\prime,y^\prime)\) or \((x,y)=(y^\prime,x^\prime)\), the exponent vanishes and hence these terms survive with probability 1. Additional off-diagonal terms could survive only if all rows \(A_i\) yield the same parity outcome for \((x,x^\prime)\) and \((y,y^\prime)\).

We now let each row \(A_i\) of \(\mathbf{A}\) be drawn independently and uniformly from \(\{0,1\}^n\). Define the two-copy moment operator averaged over both \(\mathbf{A}\) and \(\boldsymbol{\vartheta}\):
\begin{equation}
\rho^{(2)}
\;\coloneqq\;
\mathbb{E}_{{\mathbf{A}},{\boldsymbol{\vartheta}}}
\Bigl[\lvert \psi_{\mathbf{A},\boldsymbol{\vartheta}}\rangle\langle\psi_{\mathbf{A},\boldsymbol{\vartheta}}\rvert^{\otimes2}\Bigr].
\end{equation}
We say that \(\rho^{(2)}\) reproduces (or approximates) the second-moment operator of the uniform diagonal unitary distribution \(\rho_{\mathrm{ideal}}^{(2)}\) if
\begin{equation}
\|\rho^{(2)} - \rho_{\mathrm{ideal}}^{(2)}\|_1 \;\le\; \varepsilon.
\end{equation}
If they match exactly, one obtains an exact diagonal 2-design~\cite{Designs}; if they are close in trace norm, one obtains an \(\varepsilon\)-approximate diagonal 2-design~\cite{Nakata2013Diagonal}.

For any fixed quadruple \((x,x^\prime),(y,y^\prime)\) not in the identity or swap configuration, each random row \(A_i\) has a non-vanishing probability \(c<1\) of distinguishing these pairs via parity. That is, there is 
a constant probability that 
\begin{equation}
    (-1)^{A_i \cdot x} + (-1)^{A_i \cdot x^\prime} \neq (-1)^{A_i \cdot y} + (-1)^{A_i \cdot y^\prime}. 
\end{equation}
Consequently, across s independent rows, 
the probability that all rows fail to distinguish them is at most \(c^s\).

Since there are at most \(2^{4n}\) such quadruples, a union bound implies that the expected number of surviving "unwanted" off-diagonal contributions in the total operator is bounded by \(2^{4n}\,\cdot\,c^s\).
By choosing \(s \gtrsim k\,n\) for some constant \(k\), we make \(2^{4n}\,c^s\) vanish exponentially in \(n\). Hence, with high probability over choices of \(\mathbf{A}\), no unwanted off-diagonal terms survive in \(\rho_{\mathbf{A},\boldsymbol{\vartheta}}^{\otimes 2}\). In that case,
\begin{equation}
\bigl\|\rho^{(2)} - \rho_{\mathrm{ideal}}^{(2)}\bigr\|_1
\;\le\; \varepsilon,
\end{equation}
showing that the averaged ensemble indeed provides an \(\varepsilon\)-approximate diagonal 2-design when s is sufficiently large (linear in \(n\)).

Formally, let \(\Upsilon(s)\) denote the joint distribution of random architectures \(\mathbf{A}\) with s rows and random phases \(\boldsymbol{\vartheta}\). Then the resulting ensemble
\begin{equation}
\bigl\{|\psi_{\mathbf{A}, \boldsymbol{\vartheta}}\rangle : (\mathbf{A},\boldsymbol{\vartheta}) \sim \Upsilon(s)\bigr\},
\end{equation}
forms an \(\varepsilon\)-approximate diagonal 2-design once \(s \in \mathcal{O}(n)\). Since diagonal 2-designs are known to \emph{anti-concentrate} \cite{Nakata2014Generating, Hangleiter2018Anticoncentration}, it follows that the output distributions of these random Hamiltonian phase states exhibit the desired anti-concentration property~\cite{Nakata2014Generating}.
\hfill$\blacksquare$

Before proving Theorem 1, we state the following lemma which underpins the decomposition of multi-qubit diagonal interactions (i.e., multi-qubit $Z$-rotations) into sums of lower-body $Z$-interactions.

\begin{lemma}[Circuit synthesis]
Let \(\ell\) and \(k\) be integers. Then any $r$-qubit diagonal rotation of the form
\begin{equation}
    \exp \left(\mathrm{i} \frac{\ell \pi}{2^k} Z_1 Z_2 \cdots Z_r\right)
\end{equation}
can be synthesized by combining at most $k$-body diagonal (i.e., $Z$-type) interactions with appropriate phases, up to global phases that do not affect measurement outcomes.
\end{lemma}

\paragraph{Proof of Lemma 1.}
Consider the operator
\begin{equation}
    U_r(\ell, k)=\exp \left(\mathrm{i} \frac{\ell \pi}{2^k} Z^{\otimes r}\right)
\end{equation}
where \(Z^{\otimes r}=Z_1 \otimes Z_2 \otimes \cdots \otimes Z_r\). Note that
\begin{equation}
    Z=I-2|1\rangle\langle 1|,
\end{equation}
so
\begin{equation}
    Z^{\otimes r}=(I-2|1\rangle\langle 1|)^{\otimes r}.
\end{equation}
We can express this expansion via binary strings \(q \in \{0,1\}^r\). Specifically, let \((\frac{Z-I}{2})_q\) denote a tensor product of \(r\) factors where each factor is \((\frac{Z-I}{2})\) if \(q_j=1\) and $I$ otherwise (for position \(j \in \{1,\dots,r\}\). Equivalently,
\begin{equation}
    (I-2|1\rangle\langle 1|)^{\otimes r}=\sum _{q \in\{0,1\}^r}(-2)^{|q|}\left(\frac{Z-I}{2}\right)_q,
\end{equation}
where \(|q|\) is the Hamming weight of \(q\). Thus,
\begin{equation}
    U_r(\ell, k)=\exp \left(\mathrm{i} \frac{\ell \pi}{2^k} \sum _{q \in\{0,1\}^r}(-2)^{|q|}\left(\frac{Z-I}{2}\right)_q\right).
\end{equation}
To confine ourselves to at most $k$-body terms, we observe that each string $q$ of weight $|q|>k$ contributes a rotation angle that is an integer multiple of $2\pi$ (since $\ell\pi/2^k$ is the base angle and any multiple of $2\pi$ is a global phase that cancels out in quantum mechanics). Consequently, we may omit all terms corresponding to $|q|>k$. 

Rewriting each \((\frac{Z-I}{2})_q\) as a tensor product of $I$ and $Z$, we see that each surviving term of Hamming weight $|q|\leq k$ is effectively a $|q|$-body $Z$-interaction. In total, the exponent can be broken down into a linear combination of diagonal operators of body-size at most $k$. Hence, $U_r(\ell, k)$ can be decomposed into a product of multi-qubit diagonal operators, each of which involves at most $k$ qubits, up to an irrelevant global phase.

\hfill\(\blacksquare\)

\begin{remark}[Alternative circuit synthesis strategies]
A similar decomposition strategy, leveraging expansions in terms of lower-weight $Z$-type interactions, has been discussed in Refs.~\cite{Shepherd2010Binary, Rajakumar2025Polynomial}.
\end{remark}

To illustrate the above construction, let us consider a concrete example for $r$-qubit rotations of the form
\begin{equation}
    \exp \left(\mathrm{i} \frac{\ell \pi}{8} Z_1 Z_2 \cdots Z_r\right),
\end{equation}
with \(\ell \in\left\{0,1, \ldots, 2^k-1\right\}\). One finds that it can be realized by composing
into the following steps.
\begin{enumerate}
    \item Single-qubit $Z$-rotations:
    \begin{equation}
        \exp \left(\mathrm{i} \alpha Z_i\right) \quad \text { for each } i \in\{1,2, \ldots, r\}
    \end{equation}
    with 
    \begin{equation}
        \alpha = \frac{(r^2 - 5r + 6)\ell  \pi}{4}. 
    \end{equation}
    \item Two-qubit $Z$-rotations,
    \begin{equation}
        \exp \left(\mathrm{i} \beta Z_i Z_j\right) \quad \text { for all distinct } i, j,
    \end{equation}
    with 
    \begin{equation}
        \beta = \frac{(3-r)\ell  \pi}{8}. 
    \end{equation}
    \item Three-qubit $Z$ rotations 
    \begin{equation}
        \exp \left(\mathrm{i} \gamma Z_i Z_j Z_k\right) \quad \text { for all distinct } i, j,k
    \end{equation}
    with 
    \begin{equation}
        \gamma = \frac{\ell  \pi}{8}. 
    \end{equation} 
    \end{enumerate}
Details of the exact angle assignments follow from analyzing the expansion of $(Z-I)^{\otimes r}$. (or, equivalently, from the binomial expansion 
viewpoint in Lemma 1) and ensuring terms of weights 1, 2, and 3 combine to yield the correct net rotation on each subset of qubits.

\paragraph{Proof of Theorem 1.}
The measurement-driven IQP circuits $\mathcal{C}_\mathrm{MD}$ interleave fan-out staircases with parallel single-qubit diagonal gates $\exp(\mathrm{i}\vartheta_{i,j}Z_j)$ for preparing Hamiltonian phase states. Any effective multi-qubit diagonal rotation in $\mathcal{C}_\mathrm{MD}$ ultimately reduces to a product of multi-qubit diagonal unitaries of the form
\begin{equation}
    \exp \left(\mathrm{i} \frac{\ell \pi}{2^k} Z_{j_1} Z_{j_2} \cdots Z_{j_r}\right)
\end{equation}
acting on distinct qubit subsets, for some integers \(\ell\) and \(r\). 

From Lemma 1, each such $r$-qubit rotation with angle \(\ell\pi/2^k\) is equivalent (up to a global phase) to combining at most \(k\)-body diagonal $Z$-interactions with integer multiples of $2\pi$. Hence, when all angles $\boldsymbol{\vartheta}$ lie in $\{\ell\,\pi/2^k\}$, $\mathcal{C}_\mathrm{MD}$ can be fully synthesized by commuting $Z$-interactions of at most $k$ qubits each.  Because all gates remain diagonal in the computational basis, they commute pairwise.  The net effect is equivalent to a $k$-local IQP with commuting $Z$-gates of weight at most $k$.

Therefore, the entire protocol, though it may appear to be a constant-depth dynamic circuit with intermediate measurements, is unitarily equivalent to a polynomial-depth standard IQP circuit containing multi-qubit diagonal gates.  By "pulling forward" all feed-forward corrections and combining them with the single-qubit or multi-qubit rotations, we see that the final action is local with up to $k$-body terms.

\hfill\(\blacksquare\)

In the main text, we introduced \textit{Criterion~1} to rigorously characterize the statistical randomness of a binary architecture matrix $\mathbf{A}\in\mathbb{Z}_2^{s\times n}$. Here, we elaborate in detail on the three underlying conditions comprising this criterion:

\begin{figure*}[tb]
\centering
\includegraphics[width=17.75cm]{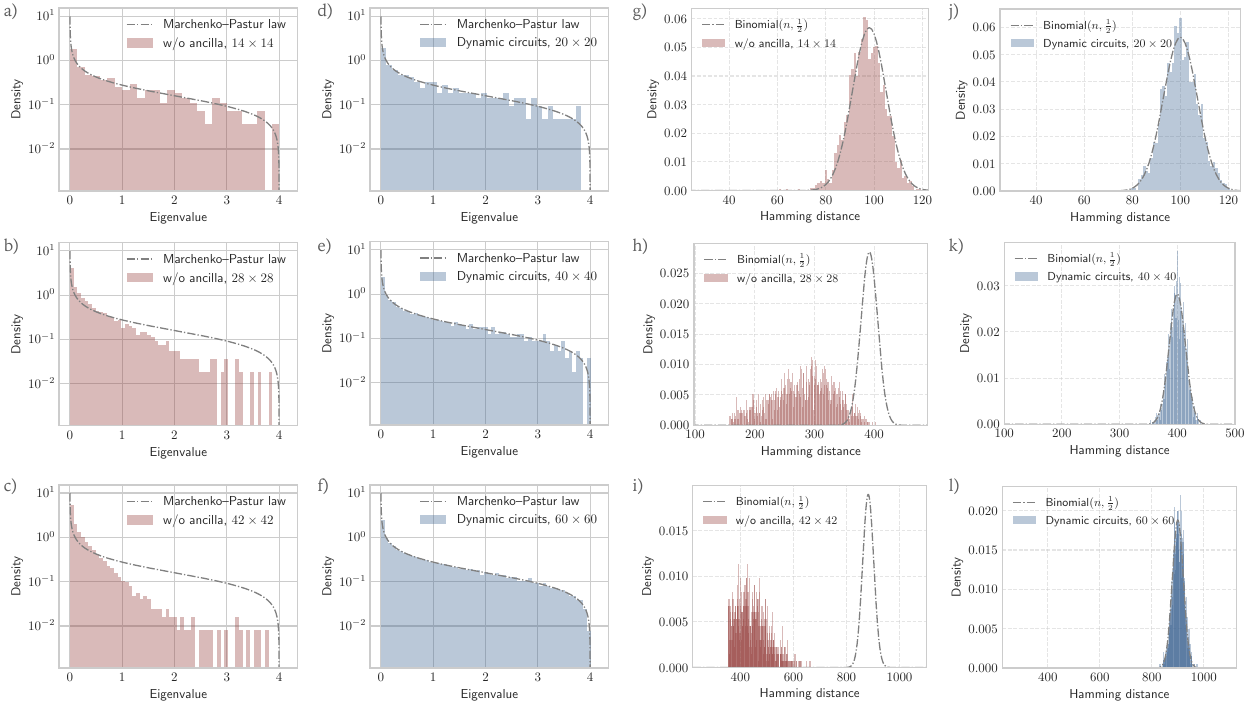}
\caption{Comparison of statistical randomness metrics under 2D lattice connectivity. Panels (a–c) depict eigenvalue distributions of standardized covariance matrices from random \(\mathrm{CX}\) circuits without auxiliary qubits at fixed depth 96 for system sizes $n=14\times14, 28\times28, 42\times42$. Panels (d–f) show eigenvalue distributions from measurement-driven randomized fan-out staircases with auxiliary qubits at fixed depth 18 (parameters: $\mathfrak{D}=3$, $\mathfrak{r}_1=2$, $\mathfrak{r}_2=1$) for system sizes $n=20\times20, 28\times28, 42\times42$. Panels (g–i) illustrate distributions of pairwise row Hamming distances for random \(\mathrm{CX}\) circuits without auxiliary qubits at depth 96, and panels (j–l) show corresponding distributions for measurement-driven circuits at depth 18, where half the qubits serve as auxiliary qubits.}
\label{fig:DetailedComparison}
\end{figure*}

\paragraph{Spectral (eigenvalue) requirement.} The standardized covariance matrix
\begin{equation}\label{eq:StandardizedCov}
Y_{\mathbf{A}}= \frac{1}{s}\bigl(2 \mathbf{A}-1\bigr)
\bigl(2 \mathbf{A}-1\bigr)^{\top}
\end{equation}
should exhibit an empirical eigenvalue distribution closely matching the \emph{Marchenko--Pastur} (MP) law~\cite{Marchenko1967Distribution}. Formally, for an aspect ratio $\gamma = n/s$ (or vice versa, depending on which dimension grows faster), the MP distribution is given by the probability density function
\begin{equation}\label{eq:MPlaw}
\begin{aligned}
\rho_{\mathrm{MP}}^{(\gamma)}(\lambda)
\;=\;&
\max\Bigl\{0,\;1 - \tfrac{1}{\gamma}\Bigr\}\,\delta(\lambda)
\\[4pt]
&+\; \frac{1}{2\pi\,\gamma\,\lambda}\,
\sqrt{
\,\bigl(\lambda - a_{-}\bigr)\,
\bigl(a_{+} - \lambda\bigr)\,
}
\\[4pt]
&\quad\text{for}\;\lambda \in [\,a_{-},\,a_{+}]\,
,
\end{aligned}
\end{equation}
where
\begin{equation}
a_{\pm}
\;=\;
\bigl(1 \,\pm\, \sqrt{\gamma}\bigr)^{2}.
\end{equation}
In the limit $s,n \to \infty$ with fixed ratio $\gamma$, random matrices of appropriate form converge almost surely to this distribution for their eigenvalue spectra.

Fig.~\ref{fig:DetailedComparison}(a–f) illustrates spectral density comparisons at fixed circuit depths across increasing system sizes, contrasting matrices generated from measurement-driven fan-out staircases with auxiliary qubits against those from local random \(\mathrm{CX}\) circuits without auxiliary qubits. As the system size grows at a fixed depth, matrices from random \(\mathrm{CX}\) circuits increasingly deviate from the ideal MP distribution, whereas eigenvalues from measurement-driven circuits remain closely aligned.

\paragraph*{Hamming (combinatorial) requirement:}
The distributions of Hamming weights—defined as the number of ones within each row $\mathbf{A}_{i,:}$ or each column $\mathbf{A}_{:,j}$—should closely follow binomial distributions $\mathrm{Binomial}(n, p)$ or $\mathrm{Binomial}(s, p)$, respectively, where $p \approx 1/2$. This implies that the entries of each row or column should behave approximately as independent random bits with equal probability.

Moreover, the pairwise Hamming distances between distinct rows or columns should similarly approximate a binomial distribution with parameter $p=1/2$. Explicitly, the pairwise Hamming distance between two distinct rows $\mathbf{A}_{i,:}$ and $\mathbf{A}_{j,:}$ is
\begin{equation}
d_{\mathrm{H}}(\mathbf{A}_{i,:},\;\mathbf{A}_{j,:})
\;=\;
\sum_{\ell = 1}^{n}
\bigl|\mathbf{A}_{i,\ell} - \mathbf{A}_{j,\ell}\bigr|,
\end{equation}
and should concentrate near $n/2$. Analogous conditions apply for columns. Satisfying these conditions ensures no systematic bias or correlations occur within the matrix entries.

Fig.~\ref{fig:DetailedComparison}(g–l) demonstrates illustrative distributions of row–row Hamming distances across various system sizes. As system size increases at fixed depth, the random \(\mathrm{CX}\) circuits without auxiliary qubits progressively deviate from the theoretical binomial distribution, while the measurement-driven circuits consistently maintain alignment with the theoretical expectation.

\paragraph*{Rank requirement over $\mathrm{GF}(2)$:} A fundamental question in random matrix theory over finite fields is: What is the probability that an $n \times n$ binary matrix, whose entries are i.i.d.\ with probability $1/2$ of being 0 or 1, has full rank $n-k$? Ref.~\cite{Kolchin1999Random} provides an explicit formula for this probability in the case of $\mathrm{GF}(2)$. Concretely, if we consider an $n \times n$ binary matrix $\mathbf{A}$ and let
\begin{equation}
    P_{n,k} \;=\; \Pr\bigl(\mathrm{rank}(\mathbf{A}) = n - k \bigr),
\end{equation}
then 
\begin{equation}\label{eq:kolchin_exact}
\begin{aligned}
P_{n,k}
=
2^{-k^2}
&\biggl(\prod_{\ell=0}^{n-k-1} \bigl(1 - 2^{-(n-\ell)}\bigr)\biggr)\\
&\cdot \biggl(\sum_{0 \le i_{1} \le \cdots \le i_{k} \le n - k}
2^{-(i_1 + \cdots + i_k)}\biggr).
\end{aligned}
\end{equation}
Here, $k$ is a fixed nonnegative integer representing the “rank deficit” from $n$.  In other words, $k=0$ corresponds to full rank, $k=1$ corresponds to rank $n-1$, etc.

When $k$ remains fixed and $n \to \infty$, $P_{n,k}$ converges to a limiting value
\begin{equation}
\begin{aligned}
\widetilde{P}_k
&\;=\;
\lim_{n\to\infty} P_{n,k}
\\&\;=\;
2^{-\,k^2}
\biggl(\,\prod_{i\ge k+1} \bigl(1 - 2^{-\,i}\bigr)\biggr)\,
\biggl(\,\prod_{\,i=1}^{\,k} \bigl(1 - 2^{-\,i}\bigr)^{-1}\biggr).
\end{aligned}
\end{equation}
Numerically, 
\begin{equation}
\begin{aligned}
\widetilde{P}_{0} \;\approx\; 0.28879,\;\;
\widetilde{P}_{1} \;\approx\; 0.57758,\;\;
\widetilde{P}_{2} \;\approx\; 0.12835,
\end{aligned}
\end{equation}
and their summation is approximately 0.99471. 

\begin{figure}[tb]
\centering
\includegraphics[width=7.5cm]{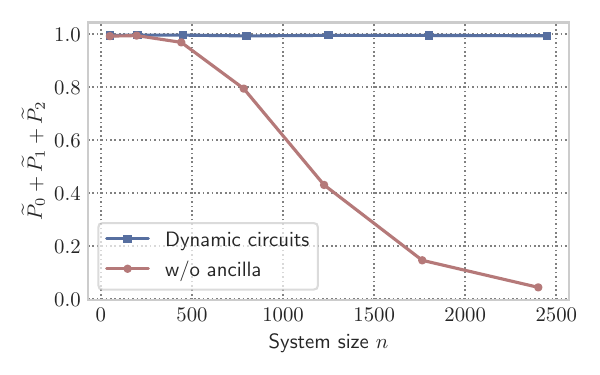}
\caption{$\widetilde{P}_0 + \widetilde{P}_1 + \widetilde{P}_2$ (the probability that a random submatrix of the $n\times n$ binary matrix has rank over $\mathrm{GF}(2)$ at least $n-2$) as a function of system size for architecture matrices $\mathbf{A}$ produced by (i) local random \(\mathrm{CX}\) circuits without auxiliary qubits (red) and (ii) measurement-driven fan-out staircases with auxiliary qubits (blue).}
\label{fig:p0p1p2}
\end{figure}

Within our setting, we require that submatrices of $\mathbf{A}$ retain high rank over $\mathrm{GF}(2)$ with large probability in order to certify algebraic randomness.  Since certain sampled matrix sizes $n$ may not be large enough to apply the strict asymptotics, we relax the criterion to demand that at least 90\% of randomly chosen submatrices achieve a rank within two of the maximum possible.  Fig.~\ref{fig:p0p1p2} compares these probabilities for architecture matrices $\mathbf{A}$ generated by (i) random \(\mathrm{CX}\) circuits without auxiliary qubits, and (ii) measurement-driven fan-out staircases with auxiliary qubits.  In particular, we plot $\widetilde{P}_0 + \widetilde{P}_1 + \widetilde{P}_2$ against system size for both methods, revealing that measurement-driven matrices maintain near-full rank across a broader range of sizes and depths.

\section{Appendix C: Geometrically local quantum circuit cost in two-dimensional lattices}

In the main text, we generalized the concept of quantum circuit cost and its entanglement-based lower bound, introduced in Ref.~\cite{Eisert2021Entangling}, to \emph{two-dimensional} (2D) lattice architectures. Quantum circuit cost quantifies the minimal resources required to implement a unitary operation by integrating the entangling capabilities of geometrically local gates. Here, we provide rigorous definitions and explicitly derive a lower bound for this cost in terms of the sum of bipartite entanglement entropies.

Consider a 2D lattice of qubits initially prepared in a product state \(|0\rangle^{\otimes n}\). Suppose a target state vector $|\psi\rangle = U|0\rangle^{\otimes n}$ is generated by a unitary $U$ composed exclusively of nearest-neighbor two-qubit interactions. Following Ref.~\cite{Eisert2021Entangling}, the 2D geometrically local circuit cost \(C^{(2D)}_{g}(U)\) is defined as the minimal total integrated interaction strength required to realize $U$. Formally, let
\begin{equation}
U = \mathcal{P}\exp\left(-\mathrm{i}\int_0^1 \mathcal{H}(s)\, ds\right)
\end{equation}
with a time-dependent Hamiltonian
\begin{equation}
\mathcal{H}(s) = \sum_{e\in\mathcal{E}} y_e(s)\, O_e,
\end{equation}
where each $O_e$ is a normalized two-qubit operator acting across lattice edge $e$, the circuit cost is then
\begin{equation}
C_g^{(2D)}(U) := \inf_{{y_e(s)}} \int_0^1 \sum_{e\in\mathcal{E}} |y_e(s)|ds.
\end{equation}

To derive the entanglement-based lower bound, consider bipartitions of the lattice defined by vertical cuts $\ell_x^{(a)}$ between columns and horizontal cuts $\ell_y^{(b)}$ between rows, each partitioning the lattice into two distinct regions. The entanglement entropy for a bipartition $\ell$ is defined as
\begin{equation}
S(\psi, \ell) := -\Tr\left[\rho_{L(\ell)} \ln \rho_{L(\ell)}\right],
\end{equation}
where $\rho_{L(\ell)}$ is the reduced density matrix on one side of the cut.

The fundamental principle underlying the lower bound is the small incremental entangling principle \cite{Marien2016Entanglement,Eisert2021Entangling}, which limits the rate of entanglement generation by local operations. Specifically, at any time $s$, the growth rate of entanglement entropy across a cut $\ell$ is bounded as
\begin{equation}
 \frac{d}{ds} S(\psi(s), \ell) \leq \eta \sum_{e\in\mathcal{E}:\, O_e \text{ crosses } \ell} |y_e(s)|
\end{equation}
with $\eta$ an $\mathcal{O}(1)$ constant dependent solely on geometry and operator norms.

Integrating this inequality over the full evolution yields
\begin{equation}
S(\psi, \ell) \leq \eta \int_0^1 \sum_{e\in\mathcal{E}:\, O_e \text{ crosses } \ell} |y_e(s)|\, ds.
\end{equation}
Since each two-qubit gate lies exactly on one nearest-neighbor cut—either horizontal or vertical—its entangling contribution is counted precisely once in either the horizontal or vertical entropy sums. Therefore, summation over all vertical and horizontal cuts gives
\begin{equation}
C_g^{(2D)}(U) \geq \frac{1}{\eta}\left(\sum_{\ell_x} S(\psi, \ell_x) + \sum_{\ell_y} S(\psi, \ell_y)\right).
\end{equation}
We define the entanglement-based quantity (scaled by a constant \(1/\eta\))
\begin{equation}
\xi(\psi) := \frac{1}{\eta}\left(\sum_{\ell_x} S(\psi, \ell_x) + \sum_{\ell_y} S(\psi, \ell_y)\right),
\end{equation}
leading directly to the inequality
\begin{equation}
\xi(\psi) \leq C_g^{(2D)}(U).
\end{equation}
Thus, the sum of bipartite entropies across all lattice cuts, appropriately scaled, constitutes a rigorous lower bound for the geometrically local quantum circuit cost in 2D lattice systems, generalizing the one-dimensional result from Ref.~\cite{Eisert2021Entangling}.

\section{Appendix D: Measurement‑based multibody XY quantum reservoir} 

For a fixed measurement-enabled architecture matrix \(\mathbf A\in\{0,1\}^{s\times n}\) (here \(s=\Theta(n)\)), define the Pauli strings
\begin{equation}
    P_{A_i}^{(\mu)}\;:=\;\bigotimes_{j=1}^{n}
\sigma^{\mu\,\mathbf A_{i,j}}_{j},
\,
\mu\in\{x,y\},\;i=1,\dots ,s.
\end{equation}
One Floquet step of the reservoir is the ordered product
\begin{equation}
    U_{\mathsf{MD}}(\tau) = \prod_{\mu = x,y} \exp\!\left(\mathrm{i}\,\tau \sum_{i=1}^{s}c^\mu_i\, P_{A_i}^{(\mu)}\right).
\end{equation}
After \(K=T/(2 \tau)\) identical steps the measurement‑driven reservoir channel reads
\begin{equation}
\mathcal R_{\mathrm{MD}}^{(T)}(\rho)=
\bigl(U_{\mathsf{MD}}(\tau)\bigr)^{K}\,
\rho\,
\bigl(U_{\mathsf{MD}}(\tau)\bigr)^{\dagger K}.
\end{equation}

The measurement‑enabled XY reservoir possesses vastly higher expressivity than any geometrically local Hamiltonian and, in principle, can extract many kinds of global information. For Theorem~2, however, it suffices to exhibit one concrete instance with only one Floquet cycle where this advantage produces a constant‑size read‑out gap.

\textit{Proof of Theorem 2.}
Let \(G=(V,E)\) be a connected graph with maximum degree \(\deg(G)=\mathcal O(1)\le \Delta\). It contains a simple
path of length at least \(n/\Delta\).  
By selecting appropriate vertices along this path, pick a triplet
\begin{equation}
  \mathcal{S}=\{i,j,k\} \subset V
\end{equation}
such that the pairwise graph distances satisfy
\begin{equation}
\operatorname{dist}(i,j),\,\operatorname{dist}(j,k),\,\operatorname{dist}(k,i)\ge \frac{n}{3\Delta}.
\end{equation}
Furthermore, choose a measurement-driven architecture matrix  \(\mathbf{A} \in \{0,1\}^{s \times n}\) so that there exists a row \(r\) with
\begin{equation}
\exists\,r:\;\mathbf{A}_{r,i}=\mathbf{A}_{r,j}=\mathbf{A}_{r,k}=1.
\label{eq:heavy-row}
\end{equation}

Let
\begin{equation}
    H_\mathcal{S}:=H_iH_jH_k,\qquad X_\mathcal{S}:=X_iX_jX_k,
\end{equation}
and set the observable \(O:=Z_iZ_jZ_k\). For \(\ell\in\{0,1\}\), define
\begin{equation}
   \ket{\psi_\ell}:=
   \frac{
      H_\mathcal{S}\ket{+}^{\otimes n}
      +(-1)^{\ell}\,\mathrm{i}\,X_\mathcal{S}H_\mathcal{S}\ket{+}^{\otimes n}
   }{\sqrt2}.
\end{equation}
Then 
\begin{equation}
    \Tr[O|\psi_0\rangle\langle\psi_0|]
 =\Tr[O|\psi_1\rangle\langle\psi_1|]=0.
\end{equation}

Now we choose the $x$‑sector couplings as
$
\widetilde{c}_r^{x}:=\pi/4
$
and, for each row $i\neq r$, we set the new coefficients
\begin{equation}
\tilde{c}_i^{x}:=
        \begin{cases}
          \frac{\varepsilon}{n}c_i^{x} & 
             \vert \mathbf{A}_i\vert_{\mathcal{S}}=1\, (\mathrm{mod}\,2),\\
          c_i^{x} & \vert \mathbf{A}_i\vert_{\mathcal{S}}=0\, (\mathrm{mod}\,2),
        \end{cases}
\end{equation}
with $\varepsilon\ll 1$, \(c_i^{x}\) being the initially sampled coefficients, \(\vert \mathbf{A}_i\vert_{\mathcal{S}}\) is the overlap of row \(A_i\) with \(\mathcal{S}\). 

For each row $i$, we choose the $y$-sector couplings
\begin{equation}
\tilde{c}_i^{y}:=
        \begin{cases}
          \frac{\varepsilon}{n}\,c_i^{y} & 
             \vert \mathbf{A}_i\vert_{\mathcal{S}}=1\, (\mathrm{mod}\,2),\,\\
          c_i^{y} & \vert \mathbf{A}_i\vert_{\mathcal{S}}=0\, (\mathrm{mod}\,2).
        \end{cases}
\end{equation}
Rows with an odd overlap \(\vert \mathbf A_i\vert_{\mathcal{S}}=1,3\) give Pauli strings that anticommute with the read‑out observable \(O=Z_iZ_jZ_k\); their couplings
are suppressed by the factor \(\varepsilon/n\) with \(0<\varepsilon\ll1\). For any such string~$\mathbf A_i$ we have the first‑order estimate
\begin{equation}
  \exp \Bigl(\frac{\mathrm i\varepsilon}{n} P^{(\mu)}_{\mathbf A_i}\Bigr)
  O
  \exp \Bigl(-\frac{\mathrm i\varepsilon}{n} P^{(\mu)}_{\mathbf A_i} \Bigr)
  =O+\mathcal O \Bigl(\frac{\varepsilon}{n}\Bigr),
\end{equation}
so their contribution to the expectation of $O$ is negligible. Rows with an even overlap \(\vert \mathbf{A}_i\vert_{\mathcal{S}}=0,2\) commute with \(O\) and
retain their sampled strength.  With these choices one Floquet period leads to 
\begin{equation}\label{eq:globalXY}
   \Tr \bigl[O\,\mathcal R_{\mathrm{MD}}^{(T)}(|\psi_0\rangle)\bigr]
      -\Tr \bigl[O\,\mathcal R_{\mathrm{MD}}^{(T)}(|\psi_1\rangle)\bigr]
      =2- \mathcal{O}\bigl(\frac{\varepsilon}{n}\bigr).
\end{equation}
Thus a single measurement‑based Floquet cycle already gives the required constant read‑out gap.

For any nearest‑neighbor Hamiltonian generating a unitary
\(V\) over the same time \(T=\mathcal O(1)\), the Lieb–Robinson bound gives
\begin{equation}
    V^{\dagger}OV=\tilde O_R + \tilde E,
\quad
\|\tilde E \| = \mathcal{O}(2^{- n}),
\end{equation}
with \(\tilde O_R\) supported within a radius‑\(R=\mathcal O(1)\) light cone.
Because the three qubits of \(S\) are mutually \(\Theta(n)\) apart,
\(\tilde O_R\) touches at most one of them; hence its expectation
coincides on \(|\psi_0\rangle\) and \(|\psi_1\rangle\). Therefore,
\begin{equation}\label{eq:localXY}
\bigl|
\Tr[O\,V|\psi_0\rangle\langle\psi_0|V^{\dagger}]
-
\Tr[O\,V|\psi_1\rangle\langle\psi_1|V^{\dagger}]
\bigr|
\le 2^{-c n}.
\end{equation}
Equations \eqref{eq:globalXY} and \eqref{eq:localXY} establish the constant–vs.–
exponential separation claimed in Theorem~2. \hfill$\blacksquare$

The same construction applies to other sector choices: for an XZ (resp.\ YZ) reservoir pick \(O=Y_iY_jY_k\) (resp.\ \(O=X_iX_jX_k\)), and choose the heavy row in a sector that anticommutes with \(O\). For the full XYZ reservoir, any choice of heavy sector with \(O\) taken in the complementary Pauli suffices. The odd/even‑overlap suppression and the Lieb–Robinson contrast carry over verbatim, yielding the same constant read‑out gap.

\section{Appendix E: Additional numerical benchmarks for the measurement‑based quantum reservoir}\label{app:benchmarks}

This section presents a series of comparative studies on the measurement-driven \emph{quantum reservoir computing} (QRC) model, analyzing its performance under different conditions and its robustness to measurement errors. In every comparison, we hold the data-generation pipeline, feature budget, and evolution-time budget constant, varying only a single factor at a time to isolate its effect.

\begin{figure}[tbh]
\centering
\includegraphics[width=7.45cm]{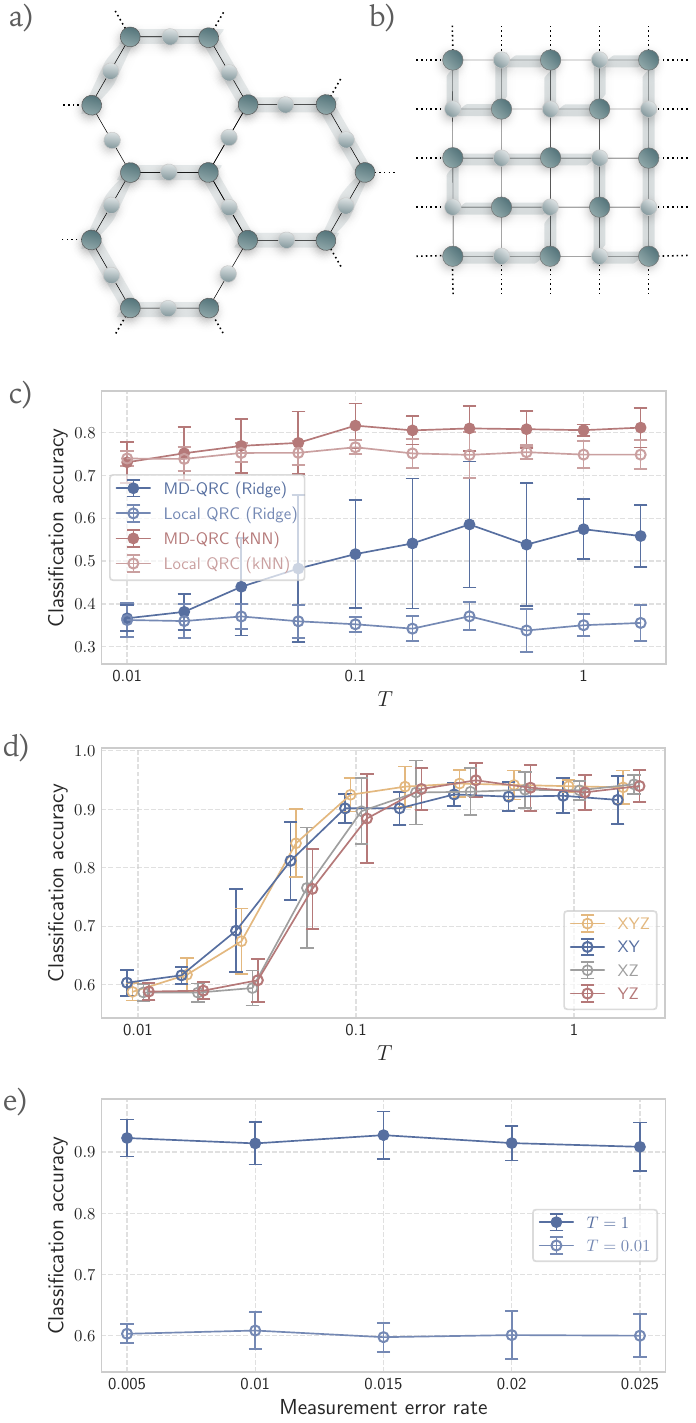}
\caption{Performance and robustness of the measurement-based QRC.
(a, b) Heavy-hex and square-lattice layouts used to define the reservoir architectures.
(c) Classifier-independent performance: The measurement-based reservoir (filled markers) consistently outperforms the local Ising baseline (open markers) for both ridge regression (blue) and $k$‑nearest neighbors (red) readouts.
(d) Pauli sector comparison: The measurement-based reservoir achieves high accuracy regardless of the chosen Pauli interaction sector $S\in\{\mathrm{XY},\mathrm{XZ},\mathrm{YZ},\mathrm{XYZ}\}$ using the SVM readout. 
(e) Robustness to measurement error: The reservoir's accuracy shows only a graceful degradation with increasing per-qubit measurement error, demonstrating feasibility with near-term hardware.
Each point is an average over 20 independent architectures (10 for each layout).}
\label{fig:S7}
\end{figure}

We generate input states from the extended SSH chain on $n=16$ system qubits, which exhibits three distinct phases (trivial, topological, and symmetry-broken). For each 
phase, we uniformly sample an eigenstate from the lowest 20 energy levels. Each sampled state is then prepared for the reservoir by applying a weak, local coherent perturbation: for every qubit \(i\), we independently draw two angles
\begin{equation}
\theta_i^{(x)}\sim\mathcal N(0,0.03),\quad
\theta_i^{(z)}\sim\mathcal N(0,0.03),
\end{equation}
and apply the rotation \(R_x\big(\theta_i^{(x)}\big)\,R_z\big(\theta_i^{(z)}\big)\).

The reservoir dynamics are governed by Hamiltonians with \(s=16\) terms, where all coupling strengths are sampled i.i.d. from \(\mathcal N(0, 1/2)\). The system undergoes ten identical Floquet cycles. We consider two hardware-motivated layouts: heavy-hex and square lattices [Figs.~\ref{fig:S7}(a,b)]. For each layout, we sample 10 independent architecture matrices (either randomized fan-out staircases or degree-matched local edge sets), meaning each data point represents an average over 20 distinct architectures. The corresponding standard deviations are shown as error bars.

The output features are the set of single-site expectation values,
\begin{equation}
\mathbf f(T)=\big(\langle\sigma_1^z\rangle,\ldots,\langle\sigma_n^z\rangle\big),
\end{equation}
which are measured at the end of each Floquet cycle up to a total evolution time \(T\). Each expectation value is estimated from 8192 shots. Unless specified otherwise, our simulation includes a symmetric per-qubit readout flip error of \(\varepsilon_{\mathrm{meas}}=5\times 10^{-3}\). The final dataset is class-balanced, and we use an identical train/test split across all methods. Features are standardized using statistics from the training set only.

We benchmark two families of reservoirs. The first, our \emph{local baseline}, is a \emph{transverse-field Ising} (TFI) model on the given hardware layout,
\begin{equation}
\mathcal H_{\mathrm{TFI}}
=\sum_{\langle i,j\rangle} J_{ij} \sigma^z_i \sigma^z_j+\sum_i h_i \sigma^x_i.
\end{equation}
The second is the \emph{measurement-based multibody reservoir}, implemented with mid-circuit measurements and Pauli-frame feed-forward as described in the main text. This method generates global interactions in a chosen Pauli sector $S\in\{\mathrm{XY},\mathrm{XZ},\mathrm{YZ},\mathrm{XYZ}\}$ from an architecture matrix $\mathbf A\in\{0,1\}^{s\times n}$:
\begin{equation}
U_{\mathsf{MD}}(\tau)=
\prod_{\mu\in S}
\exp\left(
\mathrm{i}\,\tau\sum_{i=1}^{s} c_i^{\mu}
\bigotimes_{j=1}^{n}\sigma_{j}^{\mu\,\mathbf A_{i,j}}
\right).
\end{equation}
For a fair comparison, all benchmarks match the total evolution time $T$ and the number of Hamiltonian terms.

To assess the classifier-independence of our results, we supplement the SVM analysis with two distinct models: ridge regression (a linear model) and $k$-nearest neighbors ($k=21$), a non-parametric method. Figure~\ref{fig:S7}(c) shows that the measurement-based reservoir consistently achieves higher accuracy than the local TFI baseline, regardless of the chosen classifier. The baseline's performance, in contrast, shows little improvement with increasing evolution time $T$. Since all other experimental conditions are identical, this strongly indicates that the observed advantage stems from the superior quantum feature map generated by the measurement-driven dynamics—specifically its global mixing and input-dependent nonlinearity—and is not an artifact of the classical readout method.

Figure~\ref{fig:S7}(d) tests the performance dependence on the specific Pauli sector used for the multibody interactions. We compare reservoirs generated from four different sectors $(S=\mathrm{XY},\mathrm{XZ},\mathrm{YZ},\mathrm{XYZ})$, keeping all other parameters fixed. The results show that all four choices yield quantitatively similar learning curves, achieving high classification accuracy at moderate evolution times. This indicates that the performance gain is robust and not tied to a specific interaction basis, but rather stems from the general ability to create global, multibody gates via measurement and feed-forward, as formalized in our expressivity separation (Theorem~2).

We also investigate the reservoir's robustness to measurement noise, a key factor for near-term implementation. As shown in Fig.~\ref{fig:S7}(e), we test the classification accuracy against a per-qubit readout error rate increasing from 0.5\% to 2.5\%. The performance exhibits a graceful degradation. For both longer ($T=1$) and shorter ($T=0.01$) evolution times, the drop in accuracy is minor and remains well within the statistical uncertainty from architecture sampling. This resilience to noise at levels typical of current quantum hardware underscores the practical feasibility of our approach.

In summary, all benchmarks presented here were performed under strictly controlled conditions: the input data, feature space $\{\langle\sigma_i^z\rangle\}$, shot budget, evolution time, and Hamiltonian parameter counts were held constant for each comparison. These controlled tests, combined with the analysis in the main text showing a significant performance drop when feed-forward is disabled, provide strong evidence for our central claim. They collectively demonstrate that the measurement-and-feed-forward mechanism itself—which enables the creation of global and nonlinear feature maps—is the primary driver of the observed performance advantage.


\end{document}